\newcommand*{\wn}{cm$^{-1}$}
\documentclass[aps,pra,twocolumn,superscriptaddress]{revtex4-2}
\usepackage{graphicx}
\usepackage{verbatim}
\usepackage{subfigure}
\usepackage[english]{babel}
\usepackage{verbatim}
\usepackage{multirow}
\usepackage{xcolor}
\usepackage[utf8]{inputenc}

\usepackage{dcolumn}
\usepackage{bm}
\usepackage[flushleft]{threeparttable}

\newcolumntype{.}{D{.}{.}{-1}}
\newcolumntype{d}[1]{D{.}{.}{#1}}

\newcommand*{\EF}{EF$^1\Sigma_g^+$}

\newcommand*{\GK}{GK$\,^1\Sigma_g^+$}

\newcommand*{\C}{C$\,^1\Pi_u$}
\usepackage[colorlinks,linkcolor=blue,anchorcolor=blue,citecolor=blue,]{hyperref}
\newcommand{\eqref}[1]{Eq.~(\ref{#1})}

\begin{document}

\title{Improved Ionization and Dissociation Energies of the Deuterium Molecule}

\author{J. Hussels}
\affiliation{Department of Physics and Astronomy, LaserLaB, Vrije Universiteit Amsterdam, de Boelelaan 1081, 1081 HV Amsterdam, The Netherlands}

\author{N. H\"{o}lsch}
\affiliation{Laboratorium f\"{u}r Physikalische Chemie, ETH Z\"{u}rich, 8093 Z\"{u}rich, Switzerland}

\author{C.-F. Cheng}
\affiliation{Department of Physics and Astronomy, LaserLaB, Vrije Universiteit Amsterdam, de Boelelaan 1081, 1081 HV Amsterdam, The Netherlands}
\affiliation{Department of Chemical Physics, University of Science and Technology of China, Hefei, 230026 China}

\author{E. J. Salumbides}
\affiliation{Department of Physics and Astronomy, LaserLaB, Vrije Universiteit Amsterdam, de Boelelaan 1081, 1081 HV Amsterdam, The Netherlands}

\author{H. L. Bethlem}
\affiliation{Department of Physics and Astronomy, LaserLaB, Vrije Universiteit Amsterdam, de Boelelaan 1081, 1081 HV Amsterdam, The Netherlands}

\author{K. S. E. Eikema}
\affiliation{Department of Physics and Astronomy, LaserLaB, Vrije Universiteit Amsterdam, de Boelelaan 1081, 1081 HV Amsterdam, The Netherlands}

\author{Ch. Jungen}
\affiliation{Department of Physics and Astronomy, University College London, London WC1E 6BT, United Kingdom}

\author{M. Beyer}
\affiliation{Department of Physics and Astronomy, LaserLaB, Vrije Universiteit Amsterdam, de Boelelaan 1081, 1081 HV Amsterdam, The Netherlands}

\author{F. Merkt}
\email{frederic.merkt@phys.chem.ethz.ch (F. Merkt)}
\affiliation{Laboratorium f\"{u}r Physikalische Chemie, ETH Z\"{u}rich, 8093 Z\"{u}rich, Switzerland}

\author{W. Ubachs}
\email{w.m.g.ubachs@vu.nl (W. Ubachs)}
\affiliation{Department of Physics and Astronomy, LaserLaB, Vrije Universiteit Amsterdam, de Boelelaan 1081, 1081 HV Amsterdam, The Netherlands}

\date{\today}

\begin{abstract}
The ionization energy of D$_2$ has been determined experimentally from  measurements involving two-photon Doppler-free vacuum-ultraviolet pulsed laser excitation and near-infrared continuous-wave laser excitation to yield $E_\mathrm{I}(\mathrm{D}_2)=124\,745.393\,739(26)$ \wn.
From this value, the dissociation energy of D$_2$ is deduced to be $D_0$(D$_2$) =  36\,748.362\,282(26) \wn, representing a 25-fold improvement over previous values, and found in good agreement (at $1.6\sigma$) with recent ab initio calculations of the 4-particle nonadiabatic relativistic energy and of quantum-electrodynamic corrections up to order $m\alpha^6$.
This result constitutes a test of quantum electrodynamics in the molecular domain, while a perspective is opened to determine nuclear charge radii from molecules.
\end{abstract}

\maketitle

\section{Introduction}

The hydrogen molecule and its isotopologues have become target species for testing quantum theories of molecular structure including quantum electrodynamics (QED), even to the extent of testing the Standard Model of physics in the low energy domain~\cite{Ubachs2016}.
The dissociation energy $D_0$ of the molecule is a benchmark for confronting theory and experiment and developments in both areas have mutually stimulated progress.
In the 1960s, an experimental value of $D_0$(D$_2$) was determined by photo-excitation to the $n=2$ limit in the molecule, first by
Herzberg and collaborators~\cite{Herzberg1961,Herzberg1970}, trying to verify agreement with theories in that period \cite{Kolos1960a,Wolniewicz1966}.
The results were later improved by Stoicheff and collaborators, who measured vacuum-ultraviolet laser-induced fluorescence~\cite{Balakrishnan1994}, and by Eyler and collaborators, who performed double-resonance laser excitation~\cite{Eyler1993,Zhang2004}.
Near-threshold spectral structures and the smooth onset of dissociation at $n=2$, however, formed a bottleneck for further progress.

The Zurich-Amsterdam collaboration proposed an alternative scheme for approaching the problem by targeting the ionization energy ($E_{\rm I}$) of the molecule, in which case very narrow levels in Rydberg series can be measured at extreme precision and extrapolated to their limit.
$E_{\rm I}$ is determined by stepwise laser excitation and $D_0$ is obtained via the thermochemical cycle:
\begin{displaymath}
D_\mathrm{0}(\mathrm{D}_2) = E_\mathrm{I}(\mathrm{D}_2) + E_\mathrm{I}(\mathrm{D}_2^+) - 2  E_\mathrm{I}(\mathrm{D}),
\label{Equation}
\end{displaymath}
using the accurate values for the ionization energy of the atom, $E_{\rm I}$(D), and the ionization energy of the molecular ion, $E_\mathrm{I}(\mathrm{D}_2^+)$.
A combination of experiments by our collaboration led to a much improved value of $D_0$(D$_2$) at $6.8 \times 10^{-4}$ \wn\ accuracy over a decade ago~\cite{Liu2010}.
At the same time an  improved theoretical approach led to a similarly accurate value for  $D_0$(D$_2$) in agreement with experiment \cite{Piszczatowski2009}.

This agreement prompted improved calculations of the Born-Oppenheimer (BO) potential~\cite{Pachucki2010} and leading-order effects of the nonadiabatic corrections \cite{Pachucki2015}.
A highly accurate treatment to solve the Schr\"odinger equation \cite{Pachucki2016} was developed, but progress in theory was halted by unexpected difficulties in the treatment of the relativistic corrections~\cite{Puchalski2017}.
A breakthrough was achieved through non-BO or 4-particle variational calculations, developed independently by different groups~\cite{Simmen2013,Wang2018b,Puchalski2018}.
A very accurate theoretical value for the dissociation energy of D$_2$ is now available ($D_0$(D$_2$) = 36\,748.362\,342 (26) \wn~\cite{Puchalski2019}), over an order of magnitude more accurate than the prevailing experimental value~\cite{Liu2010}.

These improvements on the theoretical side, obtained in a similar fashion also for H$_2$, represent a challenge for experiment.
Two strategies were developed to increase the experimental accuracy of the determination of the ionization and dissociation limits of the hydrogen molecules.
Firstly, a pathway through the \GK\ state, based on two-photon VUV laser excitation, was suggested~\cite{Sprecher2011} and employed, leading to greatly improved threshold values for H$_2$, both for ortho-H$_2$~\cite{Cheng2018} and para-H$_2$~\cite{Beyer2019}.
Secondly, the technique of Ramsey-comb spectroscopy utilizing frequency comb lasers for direct excitation of the \EF\ state in H$_2$ was explored~\cite{Altmann2018} and combined with narrowband laser excitation of Rydberg states to yield the most accurate value for the dissociation and ionization energies in H$_2$~\cite{Holsch2019}.

\begin{figure}[ht]
    \includegraphics[trim=0.2cm 0cm 0cm 1cm, clip=true, width=0.95\columnwidth]{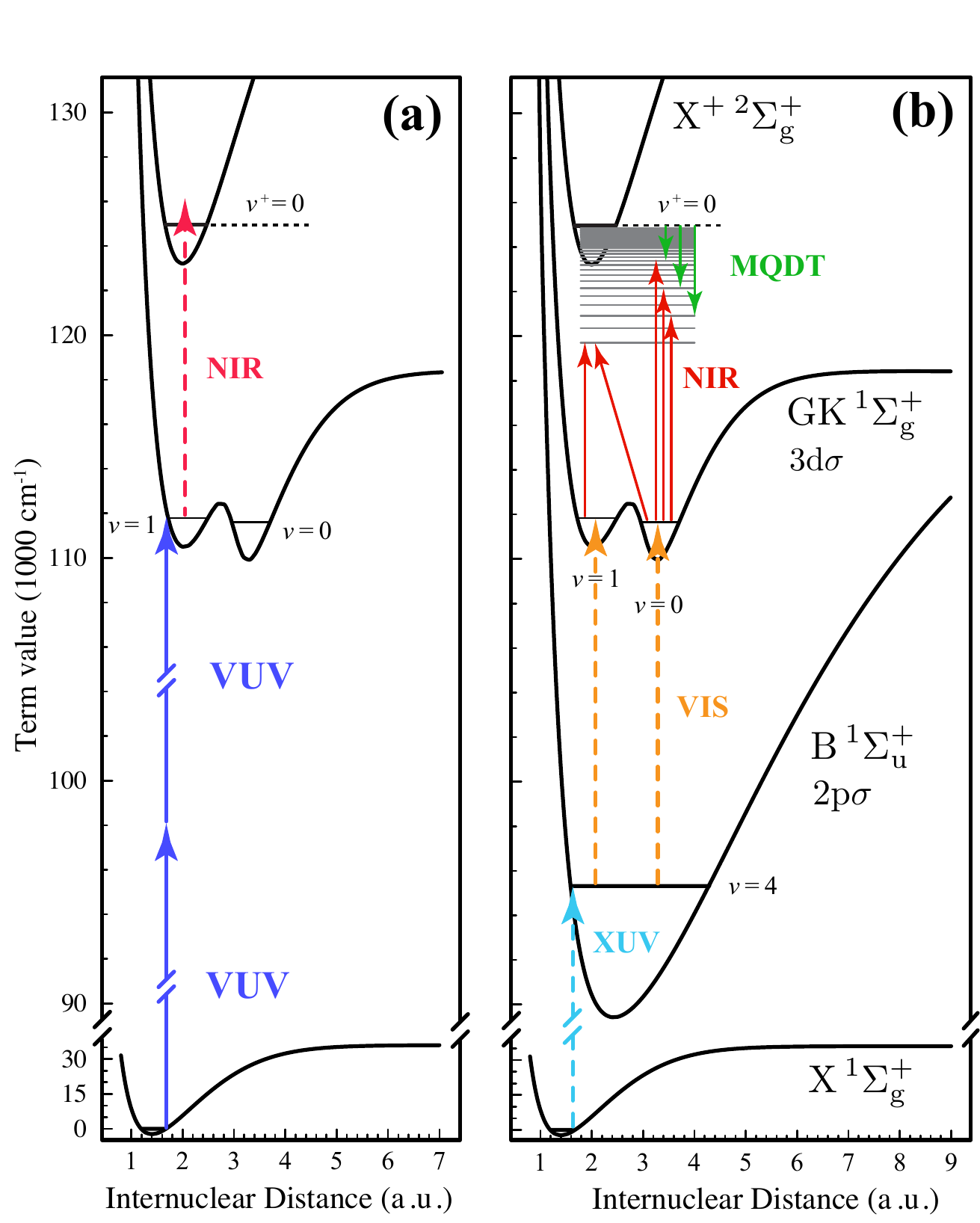}
    \caption{Potential-energy curves of the electronic states of the hydrogen molecule relevant to this study. The level positions of the Rydberg states (gray) are not to scale. (a) In Amsterdam, the {GK($v=1,N=2$)$\,\leftarrow\,$X($v=0,N=0$)} two-photon transition, indicated in dark blue, was measured to determine the term value of the GK($v=1,N=2$) state. The excited molecules were detected through ionization with near-infrared (NIR) laser radiation. (b) In Zurich, both the GK($v=1,N=2$) and GK($v=0,N=2$) levels were populated from the ground state in stepwise two-photon excitation schemes via the B($v=4,N=1$) state and their relative positions obtained by measuring transitions to the same hyperfine component of a low-$n$ Rydberg state. Starting from the long-lived GK($v=0,N=2$) state, $n$f Rydberg states belonging to series converging to the rovibrational ground state of the D$_2^+$ ion were measured for a range of $n$ values. The green arrows indicate the results from multi-channel quantum defect theory for the binding energies of the measured $n$f levels, forming an extrapolation to the ionization limit $E_{\rm I}$(D$_2$).
    \label{Fig_Levels}
    }
\end{figure}

Here, we present new results based on stepwise excitation through the high-lying \GK\ state in the D$_2$ molecule.
As in the previous studies on H$_2$, the measurements were partly carried out in the Amsterdam and Zurich laboratories and their results combined.
The excitation schemes are represented in a level diagram of the molecule in Fig.~\ref{Fig_Levels}.

\section{GK-state spectroscopy in Amsterdam}

In the Amsterdam laboratory, the \GK($v=1,J=2$) - X$\,^1\Sigma^+_g$($v=0,J=0$) energy interval in D$_2$, also referred to as the GK-X S(0) transition energy, is measured in a two-photon Doppler-free configuration.
In order to generate the required narrowband 178~nm vacuum ultraviolet (VUV) laser light, near-infrared light from a continuous-wave (CW) titanium-sapphire (Ti:Sa) laser at 714~nm is pulse amplified in a Ti:Sa oscillator-amplifier system~\cite{Hussels2021b} and harmonically upconverted using a $\beta$-BaB$_2$O$_4$ (BBO) and a  KBe$_2$BO$_3$F$_2$ (KBBF)-crystal.
The main improvement of the setup in comparison to the setup used in previous experiments~\cite{Cheng2018, Beyer2019} is the implementation of a liquid-N$_2$-cooled  valve for producing the pulsed molecular beam, which allows us to determine the molecule velocities more precisely for assessing the residual Doppler shift.
Molecules in the GK$\,^1\Sigma^+_g$(1,2) excited state are detected by selective ionization, employing autoionization resonances, excited with a single visible photon from an auxiliary pulsed dye laser.
Two measurement campaigns were performed, the first one using an autoionization resonance at high energy, $125\,899.0\,(1.0)$~\wn, the second using a resonance near the ionization threshold, at $124\,744.0\,(1.0)$~\wn.
In the former case, the ionization laser was delayed by 30~ns with respect to the VUV laser to reduce the AC-Stark effect caused by the ionization laser.
In the latter case, the AC-Stark effect was sufficiently small at zero delay, so that the VUV and visible laser pulses were overlapped.
Systematic studies were performed to estimate the contributions to the error budget by the AC-Stark effect (0.10~MHz from the visible laser and 0.24~MHz from the VUV laser inducing the two-photon transition).
For details on these AC-Stark analyses we refer to a recent PhD Thesis~\cite{Hussels2021}.

\begin{figure}[b]
    \centering
    \includegraphics[width=0.9\linewidth]{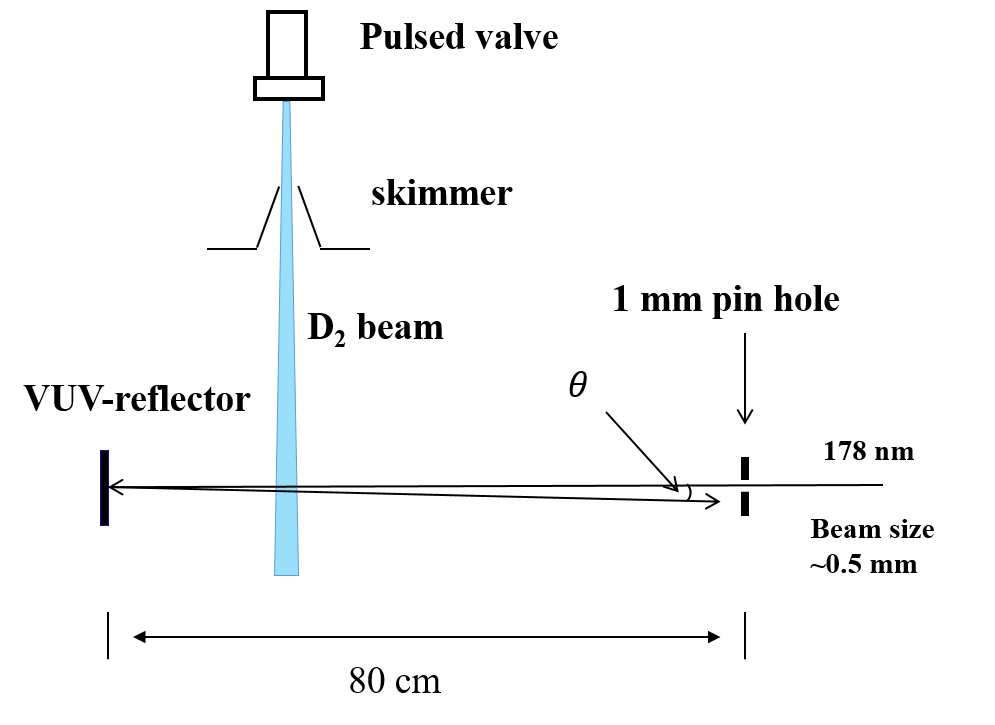}
     \caption{Schematic overview of the interaction zone of the D$_2$ molecular beam with the VUV laser beam. A small residual Doppler shift may arise when the angle $\theta$ between the incoming and reflected beam  is not exactly $180^\circ$.}
     \label{fig:dopplerangle}
\end{figure}

Even though the experiment is Doppler-free, small residual Doppler shifts persist in case of slight misalignments from exact counter-propagation of the two VUV beams crossing the molecular D$_2$ beam.
A small angle between the counter-propagating beams deviating from $180^\circ$, will give rise to a first-order Doppler shift.
To reduce this residual Doppler shift, we ascertain that the incoming and reflected VUV beams both pass through a 1 mm pinhole at a distance of 80 cm from the retro-reflecting mirror, as show in Fig.~\ref{fig:dopplerangle}.
In this way, the angle between the counter-propagating beams is constrained to less than 0.6~mrad, limiting the residual Doppler shift to 6~MHz.
This residual Doppler shift is compensated by measuring the GK$\,^1\Sigma^+_g$(1,2)$\,\leftarrow\,$X$\,^1\Sigma^+_g$(0,0) transition frequency as a function of the mean velocity of the molecular beam.
The velocity of the probed molecules was varied in a controlled manner by changing the temperature of the pulsed valve and by changing the delay time between the trigger that opens the valve and the trigger of the pulsed lasers.
The average velocity of the beam decreases from about 1700~m/s to 850~m/s upon cooling the valve from room-temperature to liquid-N$_2$ temperature (77~K), while the intensity of the molecular beam increases by about a factor of two.
Doppler-free transition frequencies are obtained from extrapolation to zero velocity for different alignments of the VUV beam.
In this analysis each measurement point is corrected for the AC-Stark effect and for the small second-order Doppler effect.

\begin{figure}[ht]
   \includegraphics[width=3.2in]{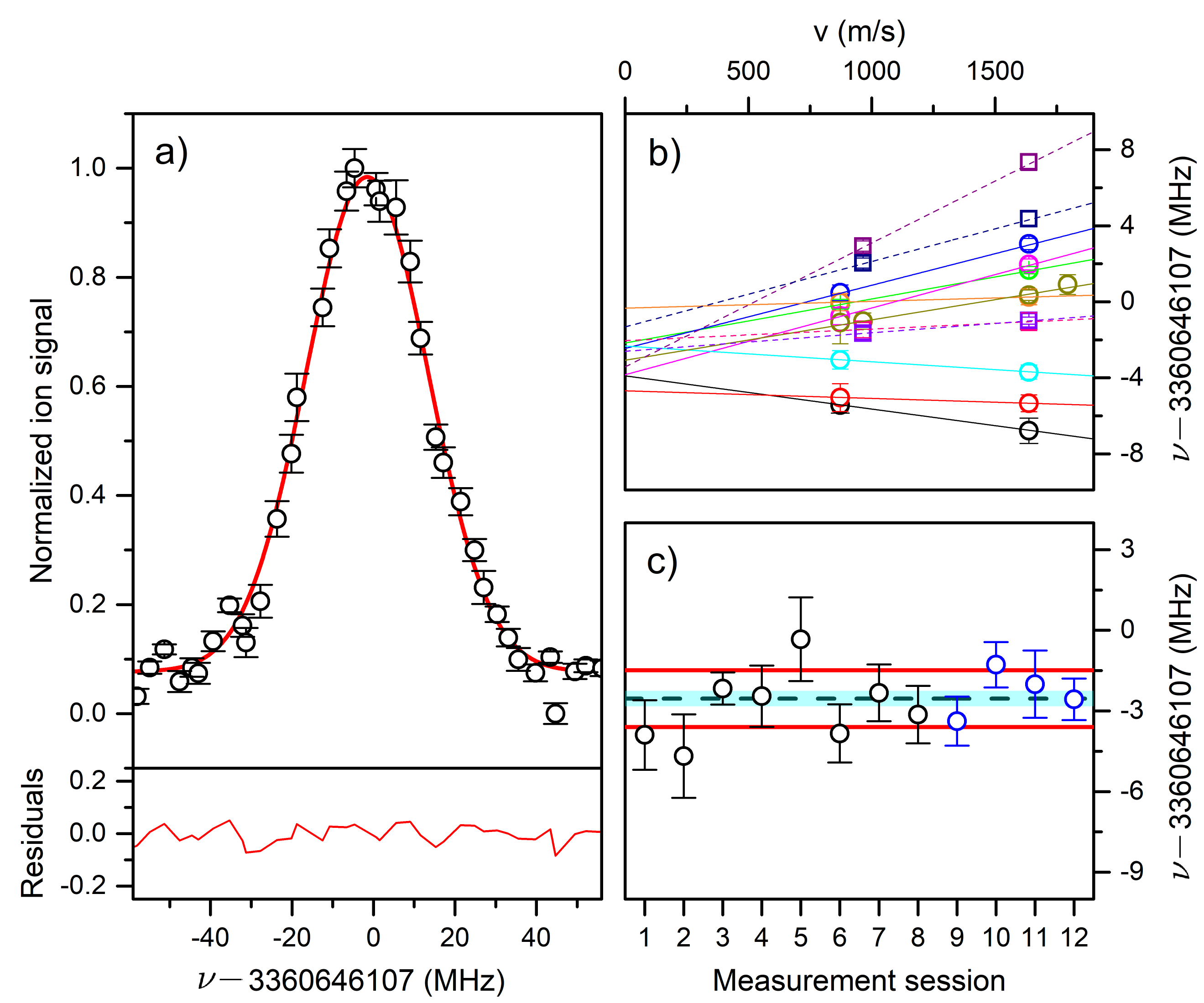}
    \caption{(a) Example of a chirp-compensated scan of the GK$\,^1\Sigma^+_g$(1,2)$\,\leftarrow\,$X$\,^1\Sigma^+_g$(0,0) two-photon transition in D$_2$, fitted with a Gaussian curve with residuals shown below.
    (b) Transition frequency as a function of central velocity of the D$_2$ beam with different colours representing different alignments of the VUV beam. Every data point results from five to ten AC-Stark and second order Doppler compensated scans. For all alignments a linear fit is used to extrapolate to a Doppler-free value. Solid lines and dashed lines result from the two separate measurement campaigns.
   (c) Extrapolated Doppler-free transition frequencies, resulting from 12 measurement sessions. Black and blue data points result from the two separate measurement campaigns. The red lines indicate the standard deviation (1~MHz). The dotted line at 3360646104.46~MHz is the average value and the blue bar is the standard error of the mean (0.35~MHz), which is the combined uncertainty due to the statistical error and the first-order Doppler shift.
    \label{Fig_VUV_Ams}
    }
\end{figure}

In Fig.~\ref{Fig_VUV_Ams} a spectrum of the GK-X S(0) line is shown. The frequency is determined from a beat-note measurement of the CW Ti:Sa seed-laser output to a frequency-comb laser, while each pulse is first chirp-compensated and thereafter analyzed for residual chirp~\cite{Hussels2021b}.
Twelve measurement sessions were carried out, each leading to a Doppler-extrapolated value at an uncertainty of 1-2 MHz (see full and dashed lines in Fig.~\ref{Fig_VUV_Ams}(b)).
This uncertainty includes statistics, chirp phenomena and residual first-order Doppler shifts, as well as the second-order Doppler effect.
Note that in the experimental configuration there is no photon-recoil effect.
Each measurement session involves a different alignment and retro-reflection of the VUV beam inside the vacuum, and can be considered to lead to independent results for which the average of the mean can be computed at an accuracy of 0.35 MHz (see Fig.~\ref{Fig_VUV_Ams}(c)).

The error budget for the VUV part of the experiments is presented in Table~\ref{tab:errD2}.
The statistical error of 0.35~MHz  includes the uncertainty of the first-order Doppler extrapolation, which itself includes the uncertainty caused by the chirp measurements.  The uncertainty in the second-order Doppler effect as well as from the hyperfine structure both result from  conservative estimates. The uncertainties of the AC-Stark shifts were different in the two measurement campaigns. The largest value of these two uncertainties is adopted.
The final uncertainty in  the GK$\,^1\Sigma^+_g$(1,2)$\,\leftarrow\,$X$\,^1\Sigma^+_g$(0,0) transition frequency of D$_2$ is 0.45~MHz.

\begin{table}[h]
    \centering
    \caption{Error budget of the measurement of the GK$\,^1\Sigma^+_g$(1,2)$\,\leftarrow\,$X$\,^1\Sigma^+_g$(0,0) transition in D$_2$.}
    \label{tab:errD2}
    \begin{tabular}{c c}
     Measured frequency    &  3360646104.46 MHz \\
     \hline
     Effect & Uncertainty \\
     \hline
      Residual 1$^\mathrm{st}$ order Doppler   & (350 kHz)$_\mathrm{stat}$\\
      2$^\mathrm{nd}$ order Doppler & 30 kHz \\
      AC-Stark ionization laser & 100 kHz \\
      AC-Stark VUV laser & 240 kHz \\
      Hyperfine structure & 100 kHz \\
      \hline
      Final frequency & 3360646104.46(45) MHz \\
      \hline
    \end{tabular}
\end{table}

The symmetric lineshape of the GK-X S(0) transition fully hides the hyperfine structure of much less than 1 MHz within a linewidth of 30 MHz.
This transition connects the center-of-gravity of the $F=0,2$ hyperfine levels of the ortho $N=0$ ground state to the GK $N=2$ excited state, with $F=0,2$, and $4$ hyperfine substates.
We include a conservative 100 kHz  estimate for the contribution of hyperfine effects associated with the unresolved VUV-transition to the error budget.
In the subsequent step, the $F=2$ component of the GK-state with $I=0$ (see below) is further excited into the Rydberg manifold; this $F=2$ level is not subject to a hyperfine interaction,
which represents an advantage of the measurement scheme possible in D$_2$ over that used in ortho-H$_2$~\cite{Cheng2018,Holsch2019}.

\section{Rydberg spectroscopy in Zurich}

The experiments carried out in  Zurich are aimed at connecting the GK$\,^1\Sigma^+_g$(1,2) level to $E_{\rm I}$(D$_2$) via measurement of transitions to high-$n$ Rydberg states. Although the final state of the GK-X S(0) transition measured in Amsterdam is located in the G inner well of the GK$\,^1\Sigma^+_g$ state, we measured transitions to long-lived Rydberg states from a rovibrational level in the K outer well, the GK(0,2) level. This level is protected from radiative decay by smaller Franck-Condon factors to the lower-lying ungerade states, in particular the \C\ state.
This effect is amplified in D$_2$ compared to H$_2$, because the twice larger reduced mass leads to lower rovibrational energies, thus increasing the tunneling barrier and enhancing the localization of the vibrational wavefunctions in the respective wells.
We measured the relevant lifetimes using the pump-probe scheme described in Ref.~\cite{Hoelsch2018} and found them to be 13.2(6)~ns and 240(40)~ns for the GK(1,2) and GK(0,2) states, respectively, with corresponding natural linewidths of $\approx12\,$MHz and below 1\,MHz, respectively.
The relative position of the GK(1,2) and GK(0,2) states was determined with an accuracy of 210~kHz by repeatedly measuring the transition frequencies from these two states to the same final state, the $(F^+=3/2, G=2)$ hyperfine component of the 50f$0_3$ Rydberg state.
For the level scheme see Fig.~\ref{Fig_Levels}.

\begin{figure}[ht]
    \includegraphics[trim=0cm 0cm 0.2cm 0cm, clip=true, width=1.0\columnwidth]{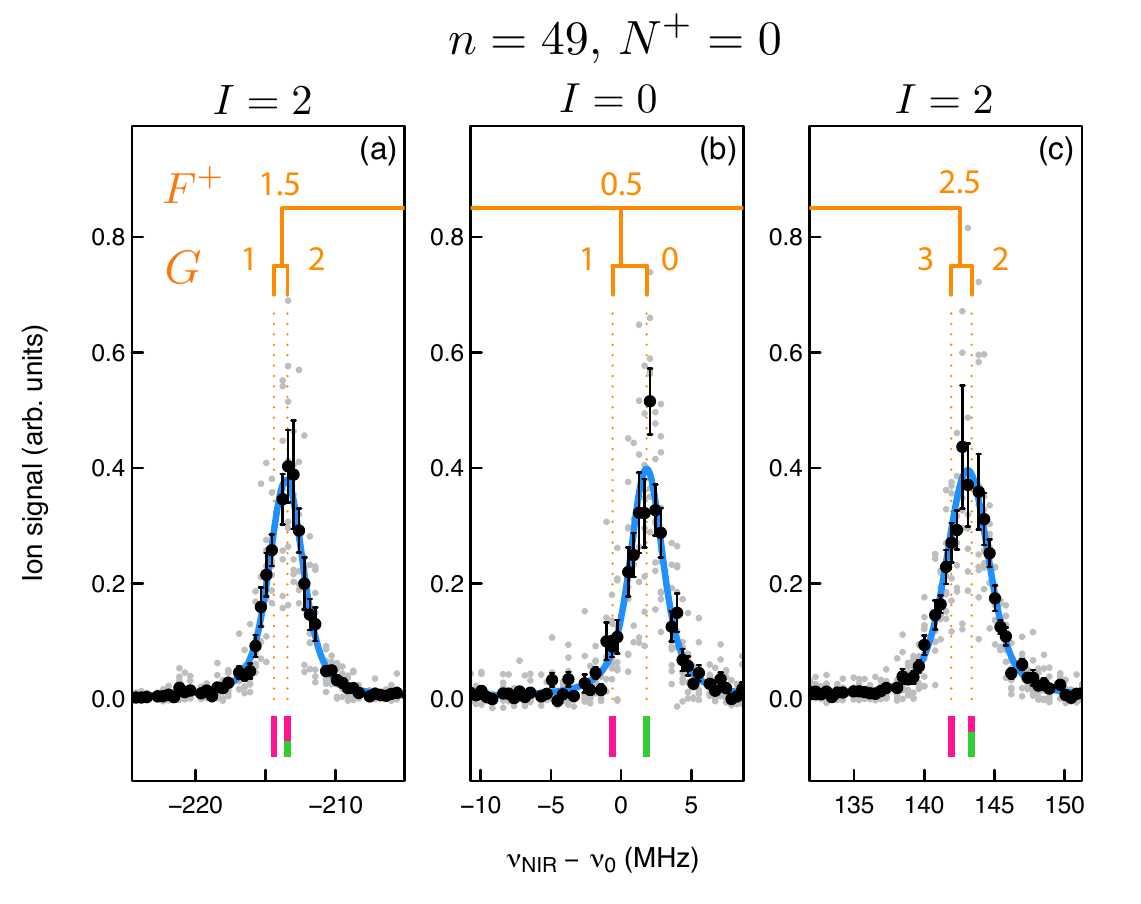}
    \caption{
    Spectra of the three $F^+=(3/2,\, 1/2,\, 5/2)$ hyperfine components of the 49f$0_3$ state of D$_2$ recorded from GK(0,2) with only the forward propagating laser beam. The normalization of the ion signal was performed for each spectrum individually and the fitted Voigt profiles are shown in blue. All three spectra were recorded without changing the geometry of the laser beam paths. $\nu_0$ was chosen as the center-of-gravity of the calculated hyperfine structure (orange) and the spectra were shifted to match the ($I=0, G=0$) component. The summed squared coefficients from all MQDT eigenvectors belonging to singlet (green) and triplet (red) basis states in Hund's case (a) are shown below the spectrum for each hyperfine component. The last coupling step to form the total angular momentum $\vec{F}=\vec{G}+\vec{\ell}$ is omitted because the splitting is not resolved on the scale of the figure.
    \label{Fig_Ryd_HFS}
    }
\end{figure}

In the absence of rotational excitation of the ion core, $N^+=0$, the fine and hyperfine structure of the molecular Rydberg state results from coupling the Rydberg electron with orbital and electron-spin angular momentum, $\vec{\ell}$ and $\vec{s}$, respectively, to the D$_2^+$ hyperfine structure. The nuclear spin allowed in the rotational ground state is $I=0,2$, resulting in the total angular momentum of the ion core given by $\vec{F}^+=\vec{G}^+=\vec{I}+\vec{S}^+=(3/2,\, 1/2,\, 5/2)$. For the rotationless molecular ion, only the Fermi-contact term in the hyperfine hamiltonian does not vanish and causes a splitting between $F^+=3/2$ and $5/2$, while leaving the $I=0$ component $F^+=1/2$ unaffected. As can be seen from Fig.~\ref{Fig_Ryd_HFS}, the observed structure in the laser spectra of the 49f0$_3 \leftarrow$\,GK(0,2) transition shows three lines, corresponding to the $F^+$ components indicated by orange bars. A multichannel-quantum-defect-theory (MQDT) calculation including spin \cite{Osterwalder2004} reveals a further splitting into a doublet for each $F^+$, which can be explained by the coupling of the Rydberg-electron spin, leading to $\vec{G} = \vec{G}^+ + \vec{s}$. The total angular momentum is then obtained through $\vec{F}=\vec{G}+\vec{\ell}$, which does not result in an observable splitting in Fig.~\ref{Fig_Ryd_HFS}.

The observed coupling scheme and intensity pattern result from the interplay of the following three interactions: i.) the Fermi-contact interaction in the ion core, ii.) the exchange interaction and iii.) the spin-orbit interaction of the Rydberg electron. Interaction (iii) is negligible, explaining the vanishing splitting of different $F$ levels for a given $G$. For states with $I=0$, interaction (i) vanishes and only the exchange interaction (ii) is present, leading to a splitting between singlet and triplet states with $S=S^+ + s$ being a good quantum number and equal to $G$. The small singlet-triplet splitting for nonpenetrating f Rydberg states was calculated to be 2.4~MHz at $n=49$ (corresponding to the splitting between $G=0,1$ for $I=0$ in Fig.~\ref{Fig_Ryd_HFS}). In terms of the quantum defects, this represents a difference of $\sim 4 \times 10^{-5}$, which is consistent with the singlet-triplet splittings observed experimentally in the H$_2$ 4f state (cf. Table II of Ref.~\cite{Uy2000}).
For $I=2$, (i) dominates over (ii), leading to $G$ states with mixed singlet-triplet character.
In Fig.~\ref{Fig_Ryd_HFS}, the stick spectra below each calculated hyperfine component indicate the corresponding singlet (green) and triplet (red) character and show good agreement with the observed spectrum, in which only the singlet character can be excited starting from the GK $^1\Sigma_g^+$ state.

Whereas the different hyperfine components of the Rydberg state cannot be resolved for $I=2$, the line
in Fig.~\ref{Fig_Ryd_HFS} (middle) arises solely from the transition to the Rydberg hyperfine component corresponding to $N^+=0, I=0, F^+=1/2, S=G=0, \ell=3, F=3$. The following discussion and analysis focuses on these states because a comparison between experimental and calculated line positions can be made at the highest precision.

\begin{figure}[ht]
    \includegraphics[trim=0cm 0cm 0.5cm 0cm, clip=true, width=1.0\columnwidth]{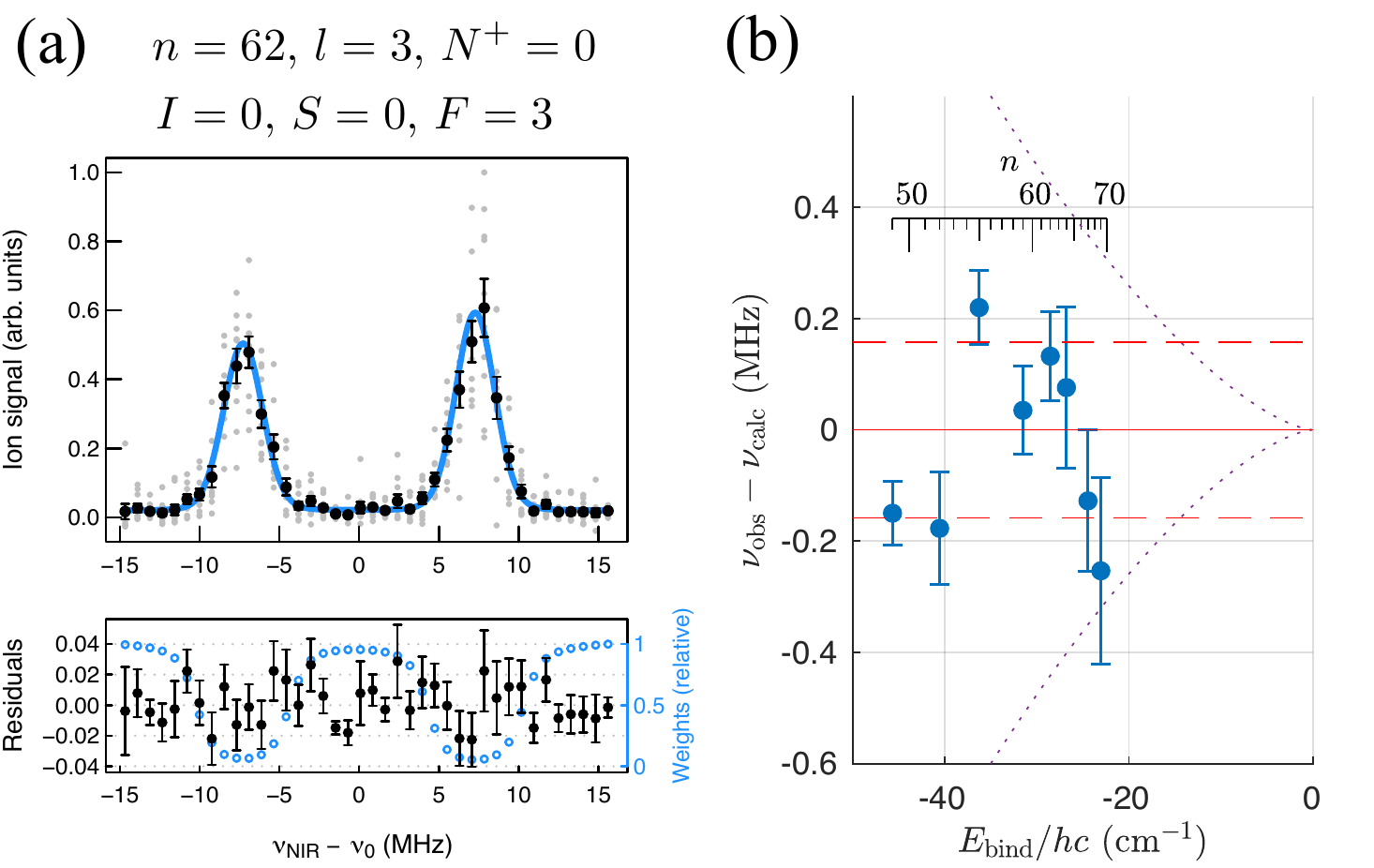}
    \caption{
    (a) Top: Determination of the Doppler-free position of the $I=0$ component from a fit of two Voigt profiles (blue) to the two Doppler components obtained with the forward and backward propagating  laser beams (grey, averages in black). Bottom: corresponding relative weights (blue) and weighted average residuals (black).
    (b) Residuals of the binding energies of the GK(0,2) state determined from $n$f$0_3 (I=0,S=0)$ Rydberg states measured for a range of selected $n$-values. The resulting statistical uncertainty of the binding energy of the GK(0,2) state is indicated by red dashed lines and the estimated uncertainty of the Rydberg-state binding energies from the MQDT treatment is indicated by the purple dotted lines.
    \label{Fig_Ryd_ETH}
    }
\end{figure}

The transitions between the GK(0,2) state and eight members of the $n$f0$_3(I=0, S=0)$  Rydberg series with $n$ values between 49 and 69 were recorded in a pulsed skimmed supersonic beam of pure D$_2$ emanating from a cryogenic pulsed valve ($T_{\rm valve}= 60$ K) using single-mode CW NIR radiation from a Ti:Sa laser.
The laser frequency was stabilized to a frequency comb which was referenced to a Rb GPS standard.
The first-order Doppler shift was cancelled by taking the average of two Doppler components generated in an optical setup in which the laser beam is retro-reflected and carefully overlapped with the forward propagating beam.
We refer to Ref.~\cite{Beyer2018b} for further details on the apparatus and measurement procedures.

A representative frequency-comb-calibrated spectrum of the resulting Doppler-doublet of the \mbox{$62\text{f}0_3(I=0,S=0)$} $\leftarrow$ GK(0,2) transition is depicted in the upper panel of Fig.~\ref{Fig_Ryd_ETH}(a), where the fitted spectrum is displayed in blue.
Stray electric fields were compensated in three dimensions, limiting the uncertainty of the transition frequencies from the DC-Stark effect to between $12\,$kHz at $n=49$ and $130\,$kHz at $n=69$.
The increase of the uncertainty with $n$ results from the $n^7$ scaling of the polarizability of Rydberg states~\cite{Gallagher1994}.
These uncertainties were added in quadrature to the respective statistical uncertainties resulting from independent sets of measurement series carried out after full realignment of the lasers to determine the total uncertainties of the Doppler-free transition frequencies.
All frequencies were corrected for the photon-recoil shift and the second-order Doppler shift.
\begin{table}[ht]
    	\centering
    	\caption{Error budget for the transition between the GK(0,2) state of D$_2$ and the 62f$0_3(I=0, S=0)$ Rydberg state, resulting from a series of independent measurements. All systematic uncertainties are the same for the other measured transitions except the uncertainty in the DC Stark shift which is $n$-dependent (see text).}
  	\label{tab:errorRydberg}
  \begin{tabular}{lrr}

    Measured frequency           & \multicolumn{2}{r}{$388\,255\,115.129(16)$~MHz}\\
    \hline
                     & Correction & Uncertainty\\
    \hline
    DC Stark shift               &         & 60~kHz\\
    AC Stark shift                          & & $\sim$5~kHz\\
    Zeeman shift                           & & $\sim$10~kHz\\
    Pressure shift                           & & $\sim$1~kHz\\
    Residual 1$^\mathrm{st}$-order Doppler shift             & & ($<$200 kHz)$_\mathrm{stat}$ \\
    2$^\mathrm{nd}$-order Doppler shift                   & +2~kHz & 0.5~kHz\\
    Line-shape model            & & 50~kHz\\
    Photon-recoil shift                 & $-$83~kHz &\\
    \hline
    Systematic uncertainty                 & & 80~kHz \\
    Final frequency                      & \multicolumn{2}{r}{$388\,255\,115.048(82)\,{\text{MHz}}$} \\
    \hline
  \end{tabular}
\end{table}

The error budget for the laser excitation of $n$f-Rydberg states from the $\text{GK}~^1\Sigma^+_g(0,2)$ state is compiled in Table~\ref{tab:errorRydberg} and includes contributions from the DC and AC-Stark shifts, Zeeman shift, pressure shift, Doppler shift and the lineshape model. The photon-recoil shift was subtracted from the observed frequencies. Our procedure to estimate all uncertainties is described in Ref.~\cite{Beyer2018b} and we only present here the procedure followed to determine the uncertainties from the DC Stark and the first-order Doppler effects, and the lineshape model.

\begin{table*}
\centering
\caption{Experimental wave numbers of the measured $n$f($N^+=0,I=0,S=0,F=3$) Rydberg states (ungerade symmetry) of ortho-D$_2$ relative to the $\text{GK}\,^1\Sigma^+_g(0,2)$ state, their statistical and combined uncertainties (in MHz), their binding energies calculated by MQDT, the predicted transition energies and the deviation between experiment and theory.}
\label{tab:linelist}
\begin{tabular}{llcc llc}
    	\hline
	$n\quad\quad$ & $\tilde{\nu}_\mathrm{obs}$ / cm$^{-1}$ & $\,\sigma_\mathrm{stat}$ / MHz$\,$ & $\,\sigma_\mathrm{tot}$ / MHz$\,$ & $E_\mathrm{bind}^\mathrm{calc}$  / cm$^{-1}$ & $\tilde{\nu}_\mathrm{calc}$ / cm$^{-1}$ & $\nu_\mathrm{obs-calc}$  / MHz \\
	\hline
$49$ &  12933.6376366 &   0.028 &   0.058 &  45.7071608 &  12933.6376416 &  $-0.149$ \\
$52$ &  12938.7605655 &   0.086 &   0.101 &  40.5842310 &  12938.7605714 &  $-0.176$ \\
$55$ &  12943.0666421 &   0.037 &   0.067 &  36.2781676 &  12943.0666348 &  $+0.220$ \\
$59$ &  12947.8198070 &   0.044 &   0.079 &  31.5249965 &  12947.8198059 &  $+0.034$ \\
$62$ &  12950.7966157 &   0.016 &   0.082 &  28.5481911 &  12950.7966113 &  $+0.133$ \\
$64$ &  12952.5524111 &   0.114 &   0.145 &  26.7923938 &  12952.5524086 &  $+0.076$ \\
$67$ &  12954.8992025 &   0.055 &   0.127 &  24.4455956 &  12954.8992068 &  $-0.128$ \\
$69$ &  12956.2956976 &   0.096 &   0.167 &  23.0490963 &  12956.2957061 &  $-0.254$ \\
 \hline
 & & & & \multicolumn{2}{r}{weighted standard error:} & 0.160
\end{tabular}
\end{table*}

\emph{DC Stark effect:} The transition to the \mbox{$62\text{f}0_3(I=0,S=0)$} state is taken as illustration because this is the transition with which the three-dimensional field compensation was performed. The combined uncertainty of the fit of the quadratic DC-Stark shift in all three spatial dimensions amounted to 60 kHz and was scaled with $n^7$ for the other states of the series, resulting in uncertainties between $12\,$kHz at $n=49$ and $130\,$kHz at $n=69$.

\emph{Lineshape model:} The statistical uncertainty for a single measurement results from the uncertainty of the nonlinear weighted fit of the lineshape of the two Doppler components (see Fig.~\ref{Fig_Ryd_ETH}a) and was found to be of the order of 50-100 kHz.
We observed Voigt profiles with linewidths (FWHM) of 2.5-3.5 MHz and in general determined the line centers to at best 1/50 of the FWHM.
We found the nonlinear fits of Voigt line profiles to be more sensitive to the starting parameters than the fits of Lorentzian line profiles which we used in our previous studies in H$_2$ \cite{Beyer2018b, Holsch2019}. This difference is most likely caused by having two fit parameters for the width instead of one. Deviations of the fitted line centers caused by fit convergence issues were observed to be of the order of 50 kHz and we cautiously took this value as additional systematic uncertainty.

\emph{Doppler effect:} The counter-propagating laser beams used to cancel the first-order Doppler shift were realigned for each independent set of measurements. This transfers the systematic error resulting from a beam misalignment, which was found to be better than 200 kHz, into a statistical error of the sample of independent measurements after recording each transition at least four and up to 14 times. The final statistical uncertainty for a measured transition was obtained as the standard error of the weighted mean of the transition frequencies from all independent measurements.

Table~\ref{tab:linelist} lists the wave numbers of all transitions $nlN^+_N(I=0,S=0)\leftarrow \text{GK}\,^1\Sigma^+_g(0,2)$ recorded in the present study. The statistical uncertainties determined from sequences of independent measurements are given as $\sigma_\mathrm{stat}$ in MHz. The systematic uncertainties were added in quadrature to obtain the final experimental uncertainties $\sigma_\mathrm{tot}$ shown in Fig.~\ref{Fig_Ryd_ETH}(b), which were used in the determination of the ionization energy of the $\text{GK}\,^1\Sigma^+_g(0,2)$ state.
Table~\ref{tab:linelist} also lists the Rydberg-state binding energies calculated by MQDT. By taking the weighted average of the sums of measured transition frequencies and MQDT binding energies, we determine the ionization energy of the GK(0,2) level to be $12979.3448024(53)\,\mathrm{cm}^{-1}$ excluding the systematic uncertainty originating from the incomplete set of quantum defects.
The second-last column gives the transition wave numbers $\tilde{\nu}_\mathrm{calc}$ obtained by subtracting the calculated Rydberg-state binding energies from this mean ionization energy. These values can be directly compared to the experimental transition wave numbers $\tilde{\nu}_\mathrm{obs}$. The differences, which are shown in the last column of Table~\ref{tab:linelist} as well as in Fig.~\ref{Fig_Ryd_ETH}(b), are all smaller than 250 kHz and do not exhibit a systematic trend.

The $l=3$ quantum-defect functions used for the MQDT calculations were extracted from available \emph{ab initio} BO potential-energy curves for low-$n$ singlet and triplet states~\cite{Silkowski2021,PachuckiPrivate}. No adjustment of the quantum defects to experimental data was performed.
We estimated the maximum possible error in the extrapolation to the ionization threshold to be 600\,kHz, including 160\,kHz originating from the residuals (root mean squared error, given by the dashed red lines in Fig.~\ref{Fig_Ryd_ETH}(b))  and the rest from the systematic uncertainty associated with the narrow range of binding energies probed experimentally. States of lower principal quantum number could not be used in this analysis because of vibrational channel interactions. As an additional verification, the extrapolation result could be confirmed within the given uncertainty using quantum defects obtained previously using a polarization model based on \emph{ab initio} data of the multipole moments of the molecular ion \cite{Jungen1989}.

\section{Discussion and conclusion}

\begin{table*}
\caption{Ionization and dissociation energies of D$_2$ from the present measurements, and comparison with previous results and with theory.
\label{D2result}}
\begin{tabular}{lc..c}
\toprule
 & Energy level interval & \multicolumn{1}{c}{Value (cm$^{-1}$)} &  \multicolumn{1}{c}{Uncertainty (MHz)}& Ref. \\
\hline
(1)& X$(v=0,N=0)$ $\to$ GK$(v=1,N=2)$          & 112099.087712(15)     & 0.45		& This work \\
(2)& GK$(v=0,N=2)$ $\to$ GK$(v=1,N=2)$         & 333.038775(7)      	& 0.21		& This work 	\\
(3)& GK$(v=0,N=2)$ $\to$ X$^+(v^+=0, N^+=0)$   & 12979.344802(20)      &   0.60		& This work \\
(4)=(1)-(2)+(3) & $E_{\rm I}$(ortho-D$_2$)        & 124745.393739(26)	    & 0.78		&	This work 	 \\
(5) &	$E_{\rm I}$(ortho-D$_2$)                  & 124745.39407(58)		& 17	    & \cite{Liu2010}  \\[4pt]
(6) &	$E_{\rm I}$(D)                            &109708.61455294(17)	& 0.005 	& \cite{Yerokhin2019} \\
(7) & $E_{\rm I}$(D$_2^+$	)                       & 131420.1976492(8)	& 0.024	    & \cite{Korobov2017, KorobovPrivate}  \\
(8)=(4)+(7)$-2\cdot$(6) & $D_0$(D$_2$)      & 36748.362282(26)		& 0.78		& This work  \\
(9) & $D_0$(D$_2$)                          & 36748.36286(68)		& 20	    & \cite{Liu2010} \\[4pt]
(10) & $D_0$(D$_2$)                          & 36748.362342(26)		& 0.80	    & Theory~\cite{Puchalski2019} \\
(8)-(10) & obs.-calc. $D_0$(D$_2$)           & -0.000060(37)	        & 1.1	\\
\hline
\end{tabular}
\end{table*}

\begin{figure}[hb]
    \includegraphics[width=1.0\columnwidth]{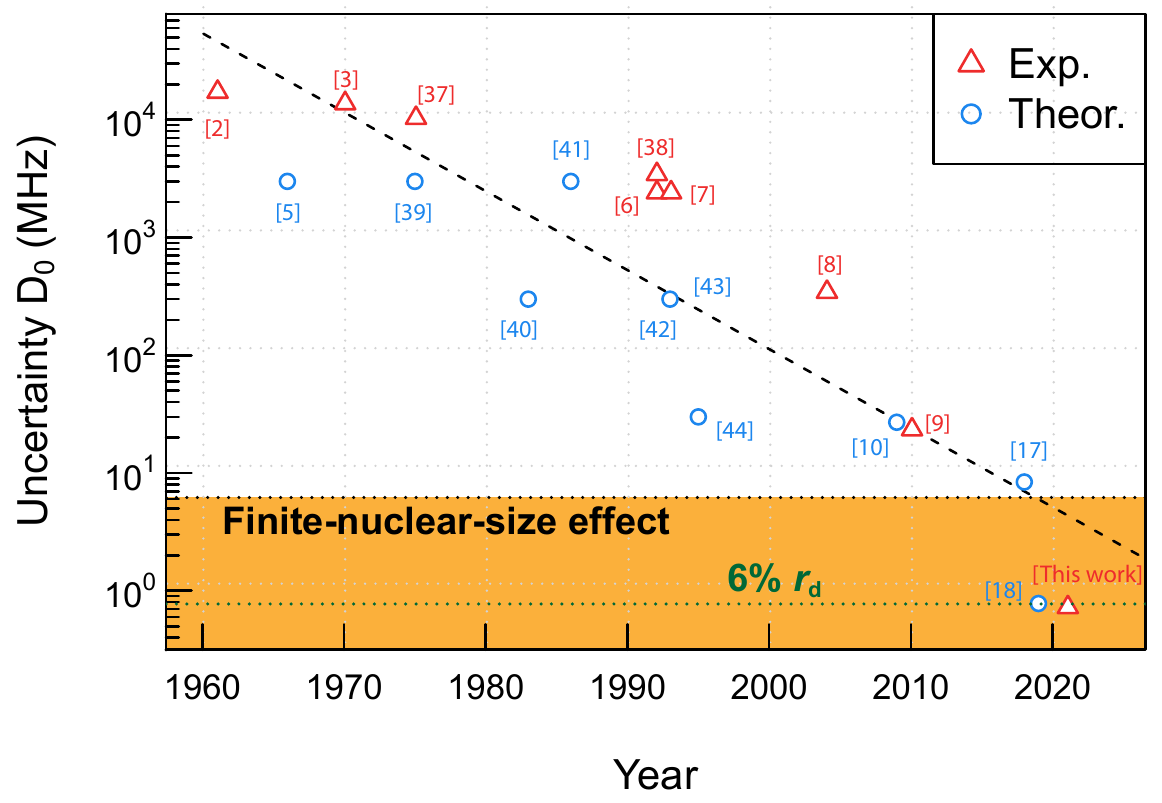}
    \caption{Comparison between experimental~\cite{Herzberg1961,Herzberg1970,Leroy1975,Jungen1992,Eyler1993,Balakrishnan1994,Zhang2004,Liu2010} and theoretical~\cite{Wolniewicz1966,Kolos1975,Wolniewicz1983,Kolos1986,Kolos1993,Wolniewicz1993,Wolniewicz1995,Piszczatowski2009,Puchalski2018,Puchalski2019} results for $D_0$(D$_2$) in a historical perspective.
    Note the improvement of $10,000$ over the plotted time span and the factor of 25 improvement over the previous experimental round.
    \label{Fig_Exp_Th}
    }
\end{figure}

The results of the present measurements and their uncertainties are compiled in Table~\ref{D2result}.
The resulting value for $D_0$(D$_2$) represents the dissociation energy of the center of gravity~\cite{Jozwiak2020} of the $N=0$ ground state.
The final outcomes are the ionization energy of the D$_2$ molecule
$E_{\rm I}$(D$_2$) = $124\,745.393\,739(26)$~\wn, and its dissociation energy $D_0$(D$_2$)= $36\,748.362\,282(26)$~\wn, which are 25 times more accurate than previous experimental results~\cite{Liu2010}.
Part of this improvement derives from the renewed calculation of $E_{\rm I}$(D$_2^+$), which was published in Ref.~\cite{Korobov2017} and subsequently updated~\cite{KorobovPrivate}. Putting this aside, the experimental improvement of $D_0$(D$_2$) would be 22-fold.
This major improvement is included in an overview of the development of this benchmark value over the past 60 years starting with the work of Herzberg~\cite{Herzberg1961,Herzberg1970}, displayed in Fig.~\ref{Fig_Exp_Th}.
A comparison is made with the development in accuracy on the theory side, demonstrating that progress on both sides goes hand-in-hand.
The experimental accuracy has improved by a factor of more than ten thousand times during this period.
A comparison with the most recent theoretical value for $D_0$(D$_2$), which is of the same accuracy, yields agreement within 1.6$\sigma$.

Precision measurements in D$_2$ are important for testing the QED framework in molecules, even though precision tests have been carried out for the H$_2$ species~\cite{Cheng2018,Beyer2019,Holsch2019}.
Inspection of the various contributions to the binding energy of the ground state of H$_2$ and D$_2$ in the most recent calculations, where nonadiabatic effects were computed in a non-BO variational approach,
shows that the various terms in the QED expansion have different contributions for the isotopologues, and hence those are tested in different combinations.
These calculations further reveal that the finite-nuclear-size (FNS) effect strongly differs between the H$_2$ and D$_2$ isotopic species.
Whereas the FNS effect in H$_2$ contributes by only $-930$~kHz to the binding energy of the molecule~\cite{Puchalski2019b}, the most accurate measurement of the binding energy is at the level of 340~kHz~\cite{Holsch2019}, so verifying the calculations at the level of 36\%.
In D$_2$ the FNS effect amounts to 6.1~MHz~\cite{Puchalski2019} because of the much larger nuclear charge radius of the deuteron (CODATA-2018 value of $r_\mathrm{d}= 2.12799 (74)$ fm~\cite{CODATA2018}).
With the current experimental precision of 780 kHz, the measured dissociation energy is sensitive to the FNS-effect below the 13\% accuracy level.
This converts to a 6\% accuracy level on $r_\mathrm{d}$, a level indicated by the orange area in Fig.~\ref{Fig_Exp_Th}.

The present accuracy for $r_\mathrm{d}$ (in CODATA-2018~\cite{CODATA2018}) is entirely based on a measurement in the muonic deuterium atom ($\mu$D)~\cite{Pohl2016}, where the overlap with the nucleus gives a FNS effect that is larger by several orders of magnitude.
It might be interesting to derive values of nuclear charge radii, without including results from muonic systems, and so pursue a derivation from measurements constrained to the first family of particles in the Standard Model of physics.
A combination of accurate results on the 1S-2S transition in atomic hydrogen~\cite{Parthey2011} and deuterium \cite{Parthey2010} and results on hydrogen and deuterium neutral molecules, including also results from the recent accurate measurements on the HD$^+$ ion~\cite{Alighanbari2020,Patra2020} and planned experiments on the H$_2^+$ ion~\cite{Schmidt2020},  open the perspective to determine accurate values of the correlated physical parameters ($r_\mathrm{d},R_{\infty}$) and ($r_\mathrm{p},R_{\infty}$), where $R_\infty$ is the Rydberg constant, from non-muonic purely electronic spectroscopy.
A comparison with results of these parameters on $r_\mathrm{p}$ and $r_\mathrm{d}$
from  $\mu$H~\cite{Pohl2010} and $\mu$D~\cite{Pohl2016} may then be interpreted as a test of lepton universality.

\vspace{0.25cm}

\section*{acknowledgments}
The authors thank M. Silkowski and K. Pachucki (Warsaw) for making available the potential-energy curves for excited states of hydrogen prior to publication.
KSEE, FM and WU acknowledge financial support form the European Research Council for ERC-Advanced grants under the European Union's Horizon 2020 research and innovation programme (No 695677, 743121, and 670168). HLB, KSEE and WU acknowledge FOM/NWO for a program grant on 'The Mysterious size of the proton'. FM acknowledges financial support from the Swiss National Science Foundation (project 200020B-200478 and synergia grant CRSII5-183579).


\begin{thebibliography}{54}%
\makeatletter
\providecommand \@ifxundefined [1]{%
 \@ifx{#1\undefined}
}%
\providecommand \@ifnum [1]{%
 \ifnum #1\expandafter \@firstoftwo
 \else \expandafter \@secondoftwo
 \fi
}%
\providecommand \@ifx [1]{%
 \ifx #1\expandafter \@firstoftwo
 \else \expandafter \@secondoftwo
 \fi
}%
\providecommand \natexlab [1]{#1}%
\providecommand \enquote  [1]{``#1''}%
\providecommand \bibnamefont  [1]{#1}%
\providecommand \bibfnamefont [1]{#1}%
\providecommand \citenamefont [1]{#1}%
\providecommand \href@noop [0]{\@secondoftwo}%
\providecommand \href [0]{\begingroup \@sanitize@url \@href}%
\providecommand \@href[1]{\@@startlink{#1}\@@href}%
\providecommand \@@href[1]{\endgroup#1\@@endlink}%
\providecommand \@sanitize@url [0]{\catcode `\\12\catcode `\$12\catcode
  `\&12\catcode `\#12\catcode `\^12\catcode `\_12\catcode `\%12\relax}%
\providecommand \@@startlink[1]{}%
\providecommand \@@endlink[0]{}%
\providecommand \url  [0]{\begingroup\@sanitize@url \@url }%
\providecommand \@url [1]{\endgroup\@href {#1}{\urlprefix }}%
\providecommand \urlprefix  [0]{URL }%
\providecommand \Eprint [0]{\href }%
\providecommand \doibase [0]{https://doi.org/}%
\providecommand \selectlanguage [0]{\@gobble}%
\providecommand \bibinfo  [0]{\@secondoftwo}%
\providecommand \bibfield  [0]{\@secondoftwo}%
\providecommand \translation [1]{[#1]}%
\providecommand \BibitemOpen [0]{}%
\providecommand \bibitemStop [0]{}%
\providecommand \bibitemNoStop [0]{.\EOS\space}%
\providecommand \EOS [0]{\spacefactor3000\relax}%
\providecommand \BibitemShut  [1]{\csname bibitem#1\endcsname}%
\let\auto@bib@innerbib\@empty
\bibitem [{\citenamefont {Ubachs}\ \emph {et~al.}(2016)\citenamefont {Ubachs},
  \citenamefont {Koelemeij}, \citenamefont {Eikema},\ and\ \citenamefont
  {Salumbides}}]{Ubachs2016}%
  \BibitemOpen
  \bibfield  {author} {\bibinfo {author} {\bibfnamefont {W.}~\bibnamefont
  {Ubachs}}, \bibinfo {author} {\bibfnamefont {J.~C.~J.}\ \bibnamefont
  {Koelemeij}}, \bibinfo {author} {\bibfnamefont {K.~S.~E.}\ \bibnamefont
  {Eikema}},\ and\ \bibinfo {author} {\bibfnamefont {E.~J.}\ \bibnamefont
  {Salumbides}},\ }\bibfield  {title} {\bibinfo {title} {Physics beyond the
  {S}tandard {M}odel from hydrogen spectroscopy},\ }\href
  {http://www.sciencedirect.com/science/article/pii/S0022285215300217}
  {\bibfield  {journal} {\bibinfo  {journal} {J. Mol. Spectrosc.}\ }\textbf
  {\bibinfo {volume} {320}},\ \bibinfo {pages} {1 } (\bibinfo {year}
  {2016})}\BibitemShut {NoStop}%
\bibitem [{\citenamefont {Herzberg}\ and\ \citenamefont
  {Monfils}(1961)}]{Herzberg1961}%
  \BibitemOpen
  \bibfield  {author} {\bibinfo {author} {\bibfnamefont {G.}~\bibnamefont
  {Herzberg}}\ and\ \bibinfo {author} {\bibfnamefont {A.}~\bibnamefont
  {Monfils}},\ }\bibfield  {title} {\bibinfo {title} {The dissociation energies
  of the {H$_2$}, {HD}, and {D$_2$} molecules},\ }\href
  {http://www.sciencedirect.com/science/article/pii/0022285261901114}
  {\bibfield  {journal} {\bibinfo  {journal} {J. Mol. Spectr.}\ }\textbf
  {\bibinfo {volume} {5}},\ \bibinfo {pages} {482} (\bibinfo {year}
  {1961})}\BibitemShut {NoStop}%
\bibitem [{\citenamefont {Herzberg}(1970)}]{Herzberg1970}%
  \BibitemOpen
  \bibfield  {author} {\bibinfo {author} {\bibfnamefont {G.}~\bibnamefont
  {Herzberg}},\ }\bibfield  {title} {\bibinfo {title} {The dissociation energy
  of the hydrogen molecule},\ }\href
  {http://www.sciencedirect.com/science/article/pii/0022285270900603}
  {\bibfield  {journal} {\bibinfo  {journal} {J. Mol. Spectrosc.}\ }\textbf
  {\bibinfo {volume} {33}},\ \bibinfo {pages} {147} (\bibinfo {year}
  {1970})}\BibitemShut {NoStop}%
\bibitem [{\citenamefont {Kolos}\ and\ \citenamefont
  {Roothaan}(1960)}]{Kolos1960a}%
  \BibitemOpen
  \bibfield  {author} {\bibinfo {author} {\bibfnamefont {W.}~\bibnamefont
  {Kolos}}\ and\ \bibinfo {author} {\bibfnamefont {C.~C.~J.}\ \bibnamefont
  {Roothaan}},\ }\bibfield  {title} {\bibinfo {title} {Correlated orbitals for
  the ground state of the hydrogen molecule},\ }\href
  {https://link.aps.org/doi/10.1103/RevModPhys.32.205} {\bibfield  {journal}
  {\bibinfo  {journal} {Rev. Mod. Phys.}\ }\textbf {\bibinfo {volume} {32}},\
  \bibinfo {pages} {205} (\bibinfo {year} {1960})}\BibitemShut {NoStop}%
\bibitem [{\citenamefont {Wolniewicz}(1966)}]{Wolniewicz1966}%
  \BibitemOpen
  \bibfield  {author} {\bibinfo {author} {\bibfnamefont {L.}~\bibnamefont
  {Wolniewicz}},\ }\bibfield  {title} {\bibinfo {title}
  {Vibrational{-}rotational study of the electronic ground state of the
  hydrogen molecule},\ }\href
  {http://scitation.aip.org/content/aip/journal/jcp/45/2/10.1063/1.1727599}
  {\bibfield  {journal} {\bibinfo  {journal} {\jcp}\ }\textbf {\bibinfo
  {volume} {45}},\ \bibinfo {pages} {515} (\bibinfo {year} {1966})}\BibitemShut
  {NoStop}%
\bibitem [{\citenamefont {Balakrishnan}\ \emph {et~al.}(1994)\citenamefont
  {Balakrishnan}, \citenamefont {Smith},\ and\ \citenamefont
  {Stoicheff}}]{Balakrishnan1994}%
  \BibitemOpen
  \bibfield  {author} {\bibinfo {author} {\bibfnamefont {A.}~\bibnamefont
  {Balakrishnan}}, \bibinfo {author} {\bibfnamefont {V.}~\bibnamefont
  {Smith}},\ and\ \bibinfo {author} {\bibfnamefont {B.~P.}\ \bibnamefont
  {Stoicheff}},\ }\bibfield  {title} {\bibinfo {title} {Dissociation energies
  of the hydrogen and deuterium molecules},\ }\href
  {http://link.aps.org/doi/10.1103/PhysRevA.49.2460} {\bibfield  {journal}
  {\bibinfo  {journal} {Phys. Rev. A}\ }\textbf {\bibinfo {volume} {49}},\
  \bibinfo {pages} {2460} (\bibinfo {year} {1994})}\BibitemShut {NoStop}%
\bibitem [{\citenamefont {Eyler}\ and\ \citenamefont
  {Melikechi}(1993)}]{Eyler1993}%
  \BibitemOpen
  \bibfield  {author} {\bibinfo {author} {\bibfnamefont {E.~E.}\ \bibnamefont
  {Eyler}}\ and\ \bibinfo {author} {\bibfnamefont {N.}~\bibnamefont
  {Melikechi}},\ }\bibfield  {title} {\bibinfo {title} {Near-threshold
  continuum structure and the dissociation energies of {H}$_{2}$, {HD}, and
  {D}$_{2}$},\ }\href {http://link.aps.org/doi/10.1103/PhysRevA.48.R18}
  {\bibfield  {journal} {\bibinfo  {journal} {Phys. Rev. A}\ }\textbf {\bibinfo
  {volume} {48}},\ \bibinfo {pages} {R18} (\bibinfo {year} {1993})}\BibitemShut
  {NoStop}%
\bibitem [{\citenamefont {Zhang}\ \emph {et~al.}(2004)\citenamefont {Zhang},
  \citenamefont {Cheng}, \citenamefont {Kim}, \citenamefont {Stanojevic},\ and\
  \citenamefont {Eyler}}]{Zhang2004}%
  \BibitemOpen
  \bibfield  {author} {\bibinfo {author} {\bibfnamefont {Y.~P.}\ \bibnamefont
  {Zhang}}, \bibinfo {author} {\bibfnamefont {C.~H.}\ \bibnamefont {Cheng}},
  \bibinfo {author} {\bibfnamefont {J.~T.}\ \bibnamefont {Kim}}, \bibinfo
  {author} {\bibfnamefont {J.}~\bibnamefont {Stanojevic}},\ and\ \bibinfo
  {author} {\bibfnamefont {E.~E.}\ \bibnamefont {Eyler}},\ }\bibfield  {title}
  {\bibinfo {title} {Dissociation energies of molecular hydrogen and the
  hydrogen molecular ion},\ }\href
  {http://link.aps.org/doi/10.1103/PhysRevLett.92.203003} {\bibfield  {journal}
  {\bibinfo  {journal} {Phys. Rev. Lett.}\ }\textbf {\bibinfo {volume} {92}},\
  \bibinfo {pages} {203003} (\bibinfo {year} {2004})}\BibitemShut {NoStop}%
\bibitem [{\citenamefont {Liu}\ \emph {et~al.}(2010)\citenamefont {Liu},
  \citenamefont {Sprecher}, \citenamefont {Jungen}, \citenamefont {Ubachs},\
  and\ \citenamefont {Merkt}}]{Liu2010}%
  \BibitemOpen
  \bibfield  {author} {\bibinfo {author} {\bibfnamefont {J.}~\bibnamefont
  {Liu}}, \bibinfo {author} {\bibfnamefont {D.}~\bibnamefont {Sprecher}},
  \bibinfo {author} {\bibfnamefont {{\mbox{Ch}}.}~\bibnamefont {Jungen}},
  \bibinfo {author} {\bibfnamefont {W.}~\bibnamefont {Ubachs}},\ and\ \bibinfo
  {author} {\bibfnamefont {F.}~\bibnamefont {Merkt}},\ }\bibfield  {title}
  {\bibinfo {title} {Determination of the ionization and dissociation energies
  of the deuterium molecule ({D}$_{2}$)},\ }\href
  {https://doi.org/10.1063/1.3374426} {\bibfield  {journal} {\bibinfo
  {journal} {\jcp}\ }\textbf {\bibinfo {volume} {132}},\ \bibinfo {pages}
  {154301} (\bibinfo {year} {2010})}\BibitemShut {NoStop}%
\bibitem [{\citenamefont {Piszczatowski}\ \emph {et~al.}(2009)\citenamefont
  {Piszczatowski}, \citenamefont {\L{}ach}, \citenamefont {Przybytek},
  \citenamefont {Komasa}, \citenamefont {Pachucki},\ and\ \citenamefont
  {Jeziorski}}]{Piszczatowski2009}%
  \BibitemOpen
  \bibfield  {author} {\bibinfo {author} {\bibfnamefont {K.}~\bibnamefont
  {Piszczatowski}}, \bibinfo {author} {\bibfnamefont {G.}~\bibnamefont
  {\L{}ach}}, \bibinfo {author} {\bibfnamefont {M.}~\bibnamefont {Przybytek}},
  \bibinfo {author} {\bibfnamefont {J.}~\bibnamefont {Komasa}}, \bibinfo
  {author} {\bibfnamefont {K.}~\bibnamefont {Pachucki}},\ and\ \bibinfo
  {author} {\bibfnamefont {B.}~\bibnamefont {Jeziorski}},\ }\bibfield  {title}
  {\bibinfo {title} {Theoretical determination of the dissociation energy of
  molecular hydrogen},\ }\href {https://doi.org/10.1021/ct900391p} {\bibfield
  {journal} {\bibinfo  {journal} {J. Chem. Theory Comput.}\ }\textbf {\bibinfo
  {volume} {5}},\ \bibinfo {pages} {3039} (\bibinfo {year} {2009})}\BibitemShut
  {NoStop}%
\bibitem [{\citenamefont {Pachucki}(2010)}]{Pachucki2010}%
  \BibitemOpen
  \bibfield  {author} {\bibinfo {author} {\bibfnamefont {K.}~\bibnamefont
  {Pachucki}},\ }\bibfield  {title} {\bibinfo {title} {Born-{O}ppenheimer
  potential for {H$_2$}},\ }\href
  {https://journals.aps.org/pra/abstract/10.1103/PhysRevA.82.032509} {\bibfield
   {journal} {\bibinfo  {journal} {Phys. Rev. A}\ }\textbf {\bibinfo {volume}
  {82}},\ \bibinfo {pages} {032509} (\bibinfo {year} {2010})}\BibitemShut
  {NoStop}%
\bibitem [{\citenamefont {Pachucki}\ and\ \citenamefont
  {Komasa}(2015)}]{Pachucki2015}%
  \BibitemOpen
  \bibfield  {author} {\bibinfo {author} {\bibfnamefont {K.}~\bibnamefont
  {Pachucki}}\ and\ \bibinfo {author} {\bibfnamefont {J.}~\bibnamefont
  {Komasa}},\ }\bibfield  {title} {\bibinfo {title} {Leading order nonadiabatic
  corrections to rovibrational levels of {H$_2$}, {D$_2$}, and {T$_2$}},\
  }\href {https://doi.org/http://dx.doi.org/10.1063/1.4927079} {\bibfield
  {journal} {\bibinfo  {journal} {J. Chem. Phys.}\ }\textbf {\bibinfo {volume}
  {143}},\ \bibinfo {pages} {034111} (\bibinfo {year} {2015})}\BibitemShut
  {NoStop}%
\bibitem [{\citenamefont {Pachucki}\ and\ \citenamefont
  {Komasa}(2016)}]{Pachucki2016}%
  \BibitemOpen
  \bibfield  {author} {\bibinfo {author} {\bibfnamefont {K.}~\bibnamefont
  {Pachucki}}\ and\ \bibinfo {author} {\bibfnamefont {J.}~\bibnamefont
  {Komasa}},\ }\bibfield  {title} {\bibinfo {title} {Schr{\"o}dinger equation
  solved for the hydrogen molecule with unprecedented accuracy},\ }\href
  {http://scitation.aip.org/content/aip/journal/jcp/144/16/10.1063/1.4948309}
  {\bibfield  {journal} {\bibinfo  {journal} {J. Chem. Phys.}\ }\textbf
  {\bibinfo {volume} {144}},\ \bibinfo {pages} {164306} (\bibinfo {year}
  {2016})}\BibitemShut {NoStop}%
\bibitem [{\citenamefont {Puchalski}\ \emph {et~al.}(2017)\citenamefont
  {Puchalski}, \citenamefont {Komasa},\ and\ \citenamefont
  {Pachucki}}]{Puchalski2017}%
  \BibitemOpen
  \bibfield  {author} {\bibinfo {author} {\bibfnamefont {M.}~\bibnamefont
  {Puchalski}}, \bibinfo {author} {\bibfnamefont {J.}~\bibnamefont {Komasa}},\
  and\ \bibinfo {author} {\bibfnamefont {K.}~\bibnamefont {Pachucki}},\
  }\bibfield  {title} {\bibinfo {title} {Relativistic corrections for the
  ground electronic state of molecular hydrogen},\ }\href
  {https://link.aps.org/doi/10.1103/PhysRevA.95.052506} {\bibfield  {journal}
  {\bibinfo  {journal} {Phys. Rev. A}\ }\textbf {\bibinfo {volume} {95}},\
  \bibinfo {pages} {052506} (\bibinfo {year} {2017})}\BibitemShut {NoStop}%
\bibitem [{\citenamefont {Simmen}\ \emph {et~al.}(2013)\citenamefont {Simmen},
  \citenamefont {Mátyus},\ and\ \citenamefont {Reiher}}]{Simmen2013}%
  \BibitemOpen
  \bibfield  {author} {\bibinfo {author} {\bibfnamefont {B.}~\bibnamefont
  {Simmen}}, \bibinfo {author} {\bibfnamefont {E.}~\bibnamefont {Mátyus}},\
  and\ \bibinfo {author} {\bibfnamefont {M.}~\bibnamefont {Reiher}},\
  }\bibfield  {title} {\bibinfo {title} {Elimination of the translational
  kinetic energy contamination in pre-{B}orn–{O}ppenheimer calculations},\
  }\href {https://doi.org/10.1080/00268976.2013.783938} {\bibfield  {journal}
  {\bibinfo  {journal} {Mol. Phys.}\ }\textbf {\bibinfo {volume} {111}},\
  \bibinfo {pages} {2086} (\bibinfo {year} {2013})}\BibitemShut {NoStop}%
\bibitem [{\citenamefont {Wang}\ and\ \citenamefont {Yan}(2018)}]{Wang2018b}%
  \BibitemOpen
  \bibfield  {author} {\bibinfo {author} {\bibfnamefont {L.}~\bibnamefont
  {Wang}}\ and\ \bibinfo {author} {\bibfnamefont {Z.-C.}\ \bibnamefont {Yan}},\
  }\bibfield  {title} {\bibinfo {title} {Relativistic corrections to the ground
  states of {HD} and {D}$_2$ calculated without using the {Born}-{Oppenheimer}
  approximation},\ }\href {https://doi.org/10.1039/C8CP04586K} {\bibfield
  {journal} {\bibinfo  {journal} {Phys. Chem. Chem. Phys.}\ }\textbf {\bibinfo
  {volume} {20}},\ \bibinfo {pages} {23948} (\bibinfo {year}
  {2018})}\BibitemShut {NoStop}%
\bibitem [{\citenamefont {Puchalski}\ \emph {et~al.}(2018)\citenamefont
  {Puchalski}, \citenamefont {Spyszkiewicz}, \citenamefont {Komasa},\ and\
  \citenamefont {Pachucki}}]{Puchalski2018}%
  \BibitemOpen
  \bibfield  {author} {\bibinfo {author} {\bibfnamefont {M.}~\bibnamefont
  {Puchalski}}, \bibinfo {author} {\bibfnamefont {A.}~\bibnamefont
  {Spyszkiewicz}}, \bibinfo {author} {\bibfnamefont {J.}~\bibnamefont
  {Komasa}},\ and\ \bibinfo {author} {\bibfnamefont {K.}~\bibnamefont
  {Pachucki}},\ }\bibfield  {title} {\bibinfo {title} {Nonadiabatic
  relativistic correction to the dissociation energy of {H}$_2$, {D}$_2$, and
  {HD}},\ }\href
  {https://journals.aps.org/prl/abstract/10.1103/PhysRevLett.121.073001}
  {\bibfield  {journal} {\bibinfo  {journal} {Phys. Rev. Lett.}\ }\textbf
  {\bibinfo {volume} {121}},\ \bibinfo {pages} {073001} (\bibinfo {year}
  {2018})}\BibitemShut {NoStop}%
\bibitem [{\citenamefont {Puchalski}\ \emph
  {et~al.}(2019{\natexlab{a}})\citenamefont {Puchalski}, \citenamefont
  {Komasa}, \citenamefont {Spyszkiewicz},\ and\ \citenamefont
  {Pachucki}}]{Puchalski2019}%
  \BibitemOpen
  \bibfield  {author} {\bibinfo {author} {\bibfnamefont {M.}~\bibnamefont
  {Puchalski}}, \bibinfo {author} {\bibfnamefont {J.}~\bibnamefont {Komasa}},
  \bibinfo {author} {\bibfnamefont {A.}~\bibnamefont {Spyszkiewicz}},\ and\
  \bibinfo {author} {\bibfnamefont {K.}~\bibnamefont {Pachucki}},\ }\bibfield
  {title} {\bibinfo {title} {Dissociation energy of molecular hydrogen
  isotopologues},\ }\href
  {https://link.aps.org/doi/10.1103/PhysRevA.100.020503} {\bibfield  {journal}
  {\bibinfo  {journal} {Phys. Rev. A}\ }\textbf {\bibinfo {volume} {100}},\
  \bibinfo {pages} {020503} (\bibinfo {year} {2019}{\natexlab{a}})}\BibitemShut
  {NoStop}%
\bibitem [{\citenamefont {Sprecher}\ \emph {et~al.}(2011)\citenamefont
  {Sprecher}, \citenamefont {Jungen}, \citenamefont {Ubachs},\ and\
  \citenamefont {Merkt}}]{Sprecher2011}%
  \BibitemOpen
  \bibfield  {author} {\bibinfo {author} {\bibfnamefont {D.}~\bibnamefont
  {Sprecher}}, \bibinfo {author} {\bibfnamefont {{\mbox{Ch}}.}~\bibnamefont
  {Jungen}}, \bibinfo {author} {\bibfnamefont {W.}~\bibnamefont {Ubachs}},\
  and\ \bibinfo {author} {\bibfnamefont {F.}~\bibnamefont {Merkt}},\ }\bibfield
   {title} {\bibinfo {title} {Towards measuring the ionisation and dissociation
  energies of molecular hydrogen with sub-{MHz} accuracy},\ }\href
  {http://dx.doi.org/10.1039/C0FD00035C} {\bibfield  {journal} {\bibinfo
  {journal} {Farad. Discuss.}\ }\textbf {\bibinfo {volume} {150}},\ \bibinfo
  {pages} {51} (\bibinfo {year} {2011})}\BibitemShut {NoStop}%
\bibitem [{\citenamefont {Cheng}\ \emph {et~al.}(2018)\citenamefont {Cheng},
  \citenamefont {Hussels}, \citenamefont {Niu}, \citenamefont {Bethlem},
  \citenamefont {Eikema}, \citenamefont {Salumbides}, \citenamefont {Ubachs},
  \citenamefont {Beyer}, \citenamefont {H\"{o}lsch}, \citenamefont {Agner},
  \citenamefont {Merkt}, \citenamefont {Tao}, \citenamefont {Hu},\ and\
  \citenamefont {Jungen}}]{Cheng2018}%
  \BibitemOpen
  \bibfield  {author} {\bibinfo {author} {\bibfnamefont {C.-F.}\ \bibnamefont
  {Cheng}}, \bibinfo {author} {\bibfnamefont {J.}~\bibnamefont {Hussels}},
  \bibinfo {author} {\bibfnamefont {M.}~\bibnamefont {Niu}}, \bibinfo {author}
  {\bibfnamefont {H.~L.}\ \bibnamefont {Bethlem}}, \bibinfo {author}
  {\bibfnamefont {K.~S.~E.}\ \bibnamefont {Eikema}}, \bibinfo {author}
  {\bibfnamefont {E.~J.}\ \bibnamefont {Salumbides}}, \bibinfo {author}
  {\bibfnamefont {W.}~\bibnamefont {Ubachs}}, \bibinfo {author} {\bibfnamefont
  {M.}~\bibnamefont {Beyer}}, \bibinfo {author} {\bibfnamefont
  {N.}~\bibnamefont {H\"{o}lsch}}, \bibinfo {author} {\bibfnamefont {J.~A.}\
  \bibnamefont {Agner}}, \bibinfo {author} {\bibfnamefont {F.}~\bibnamefont
  {Merkt}}, \bibinfo {author} {\bibfnamefont {L.-G.}\ \bibnamefont {Tao}},
  \bibinfo {author} {\bibfnamefont {S.-M.}\ \bibnamefont {Hu}},\ and\ \bibinfo
  {author} {\bibfnamefont {{\mbox{Ch}}.}~\bibnamefont {Jungen}},\ }\bibfield
  {title} {\bibinfo {title} {Dissociation energy of the hydrogen molecule at
  $10^{-9}$ accuracy},\ }\href
  {https://journals.aps.org/prl/abstract/10.1103/PhysRevLett.121.013001}
  {\bibfield  {journal} {\bibinfo  {journal} {Phys. Rev. Lett.}\ }\textbf
  {\bibinfo {volume} {121}},\ \bibinfo {pages} {013001} (\bibinfo {year}
  {2018})}\BibitemShut {NoStop}%
\bibitem [{\citenamefont {Beyer}\ \emph {et~al.}(2019)\citenamefont {Beyer},
  \citenamefont {H\"olsch}, \citenamefont {Hussels}, \citenamefont {Cheng},
  \citenamefont {Salumbides}, \citenamefont {Eikema}, \citenamefont {Ubachs},
  \citenamefont {Jungen},\ and\ \citenamefont {Merkt}}]{Beyer2019}%
  \BibitemOpen
  \bibfield  {author} {\bibinfo {author} {\bibfnamefont {M.}~\bibnamefont
  {Beyer}}, \bibinfo {author} {\bibfnamefont {N.}~\bibnamefont {H\"olsch}},
  \bibinfo {author} {\bibfnamefont {J.}~\bibnamefont {Hussels}}, \bibinfo
  {author} {\bibfnamefont {C.-F.}\ \bibnamefont {Cheng}}, \bibinfo {author}
  {\bibfnamefont {E.~J.}\ \bibnamefont {Salumbides}}, \bibinfo {author}
  {\bibfnamefont {K.~S.~E.}\ \bibnamefont {Eikema}}, \bibinfo {author}
  {\bibfnamefont {W.}~\bibnamefont {Ubachs}}, \bibinfo {author} {\bibfnamefont
  {{\mbox{Ch}}.}~\bibnamefont {Jungen}},\ and\ \bibinfo {author} {\bibfnamefont
  {F.}~\bibnamefont {Merkt}},\ }\bibfield  {title} {\bibinfo {title}
  {Determination of the interval between the ground states of para- and
  ortho-{${\mathrm{H}}_{2}$}},\ }\href
  {https://doi.org/10.1103/PhysRevLett.123.163002} {\bibfield  {journal}
  {\bibinfo  {journal} {Phys. Rev. Lett.}\ }\textbf {\bibinfo {volume} {123}},\
  \bibinfo {pages} {163002} (\bibinfo {year} {2019})}\BibitemShut {NoStop}%
\bibitem [{\citenamefont {Altmann}\ \emph {et~al.}(2018)\citenamefont
  {Altmann}, \citenamefont {Dreissen}, \citenamefont {Salumbides},
  \citenamefont {Ubachs},\ and\ \citenamefont {Eikema}}]{Altmann2018}%
  \BibitemOpen
  \bibfield  {author} {\bibinfo {author} {\bibfnamefont {R.~K.}\ \bibnamefont
  {Altmann}}, \bibinfo {author} {\bibfnamefont {L.~S.}\ \bibnamefont
  {Dreissen}}, \bibinfo {author} {\bibfnamefont {E.~J.}\ \bibnamefont
  {Salumbides}}, \bibinfo {author} {\bibfnamefont {W.}~\bibnamefont {Ubachs}},\
  and\ \bibinfo {author} {\bibfnamefont {K.~S.~E.}\ \bibnamefont {Eikema}},\
  }\bibfield  {title} {\bibinfo {title} {Deep-ultraviolet frequency metrology
  of {H}$_{2}$ for tests of molecular quantum theory},\ }\href
  {https://link.aps.org/doi/10.1103/PhysRevLett.120.043204} {\bibfield
  {journal} {\bibinfo  {journal} {Phys. Rev. Lett.}\ }\textbf {\bibinfo
  {volume} {120}},\ \bibinfo {pages} {043204} (\bibinfo {year}
  {2018})}\BibitemShut {NoStop}%
\bibitem [{\citenamefont {H\"{o}lsch}\ \emph {et~al.}(2019)\citenamefont
  {H\"{o}lsch}, \citenamefont {Beyer}, \citenamefont {Salumbides},
  \citenamefont {Eikema}, \citenamefont {Ubachs}, \citenamefont {Jungen},\ and\
  \citenamefont {Merkt}}]{Holsch2019}%
  \BibitemOpen
  \bibfield  {author} {\bibinfo {author} {\bibfnamefont {N.}~\bibnamefont
  {H\"{o}lsch}}, \bibinfo {author} {\bibfnamefont {M.}~\bibnamefont {Beyer}},
  \bibinfo {author} {\bibfnamefont {E.~J.}\ \bibnamefont {Salumbides}},
  \bibinfo {author} {\bibfnamefont {K.~S.~E.}\ \bibnamefont {Eikema}}, \bibinfo
  {author} {\bibfnamefont {W.}~\bibnamefont {Ubachs}}, \bibinfo {author}
  {\bibfnamefont {{\mbox{Ch}}.}~\bibnamefont {Jungen}},\ and\ \bibinfo {author}
  {\bibfnamefont {F.}~\bibnamefont {Merkt}},\ }\bibfield  {title} {\bibinfo
  {title} {Benchmarking theory with an improved measurement of the ionization
  and dissociation energies of {H}$_2$},\ }\href
  {https://doi.org/10.1103/PhysRevLett.122.103002} {\bibfield  {journal}
  {\bibinfo  {journal} {Phys. Rev. Lett.}\ }\textbf {\bibinfo {volume} {122}},\
  \bibinfo {pages} {103002} (\bibinfo {year} {2019})}\BibitemShut {NoStop}%
\bibitem [{\citenamefont {Hussels}\ \emph {et~al.}(2020)\citenamefont
  {Hussels}, \citenamefont {Cheng}, \citenamefont {Salumbides},\ and\
  \citenamefont {Ubachs}}]{Hussels2021b}%
  \BibitemOpen
  \bibfield  {author} {\bibinfo {author} {\bibfnamefont {J.}~\bibnamefont
  {Hussels}}, \bibinfo {author} {\bibfnamefont {C.}~\bibnamefont {Cheng}},
  \bibinfo {author} {\bibfnamefont {E.~J.}\ \bibnamefont {Salumbides}},\ and\
  \bibinfo {author} {\bibfnamefont {W.}~\bibnamefont {Ubachs}},\ }\bibfield
  {title} {\bibinfo {title} {Chirp-compensated pulsed titanium--sapphire laser
  system for precision spectroscopy},\ }\href
  {http://ol.osa.org/abstract.cfm?URI=ol-45-21-5909} {\bibfield  {journal}
  {\bibinfo  {journal} {Opt. Lett.}\ }\textbf {\bibinfo {volume} {45}},\
  \bibinfo {pages} {5909} (\bibinfo {year} {2020})}\BibitemShut {NoStop}%
\bibitem [{\citenamefont {Hussels}(2021)}]{Hussels2021}%
  \BibitemOpen
  \bibfield  {author} {\bibinfo {author} {\bibfnamefont {J.}~\bibnamefont
  {Hussels}},\ }\emph {\bibinfo {title} {Improved determination of the
  dissociation energy of {H}$_2$, {HD} and {D}$_2$}},\ \href@noop {} {Ph.D.
  thesis},\ \bibinfo  {school} {Vrije Universiteit Amsterdam} (\bibinfo {year}
  {2021})\BibitemShut {NoStop}%
\bibitem [{\citenamefont {H{\"o}lsch}\ \emph {et~al.}(2018)\citenamefont
  {H{\"o}lsch}, \citenamefont {Beyer},\ and\ \citenamefont
  {Merkt}}]{Hoelsch2018}%
  \BibitemOpen
  \bibfield  {author} {\bibinfo {author} {\bibfnamefont {N.}~\bibnamefont
  {H{\"o}lsch}}, \bibinfo {author} {\bibfnamefont {M.}~\bibnamefont {Beyer}},\
  and\ \bibinfo {author} {\bibfnamefont {F.}~\bibnamefont {Merkt}},\ }\bibfield
   {title} {\bibinfo {title} {Nonadiabatic effects on the positions and
  lifetimes of the low-lying rovibrational levels of the {GK}{$^1\Sigma_g^+$}
  and {H$^1\Sigma_g^+$} states of {H}$_2$},\ }\href
  {https://doi.org/10.1039/C8CP05233F} {\bibfield  {journal} {\bibinfo
  {journal} {Phys. Chem. Chem. Phys.}\ }\textbf {\bibinfo {volume} {20}},\
  \bibinfo {pages} {26837} (\bibinfo {year} {2018})}\BibitemShut {NoStop}%
\bibitem [{\citenamefont {Osterwalder}\ \emph {et~al.}(2004)\citenamefont
  {Osterwalder}, \citenamefont {{W\"{u}est}}, \citenamefont {Merkt},\ and\
  \citenamefont {Jungen}}]{Osterwalder2004}%
  \BibitemOpen
  \bibfield  {author} {\bibinfo {author} {\bibfnamefont {A.}~\bibnamefont
  {Osterwalder}}, \bibinfo {author} {\bibfnamefont {A.}~\bibnamefont
  {{W\"{u}est}}}, \bibinfo {author} {\bibfnamefont {F.}~\bibnamefont {Merkt}},\
  and\ \bibinfo {author} {\bibfnamefont {{\mbox{Ch}}.}~\bibnamefont {Jungen}},\
  }\bibfield  {title} {\bibinfo {title} {High-resolution millimeter wave
  spectroscopy and multichannel quantum defect theory of the hyperfine
  structure in high {R}ydberg states of molecular hydrogen {H$_2$}},\ }\href
  {http://scitation.aip.org/content/aip/journal/jcp/121/23/10.1063/1.1792596}
  {\bibfield  {journal} {\bibinfo  {journal} {\jcp}\ }\textbf {\bibinfo
  {volume} {121}},\ \bibinfo {pages} {11810} (\bibinfo {year}
  {2004})}\BibitemShut {NoStop}%
\bibitem [{\citenamefont {Uy}\ \emph {et~al.}(2000)\citenamefont {Uy},
  \citenamefont {Gabrys}, \citenamefont {Oka}, \citenamefont {Cotterell},
  \citenamefont {Stickland}, \citenamefont {Jungen},\ and\ \citenamefont
  {Wüest}}]{Uy2000}%
  \BibitemOpen
  \bibfield  {author} {\bibinfo {author} {\bibfnamefont {D.}~\bibnamefont
  {Uy}}, \bibinfo {author} {\bibfnamefont {C.~M.}\ \bibnamefont {Gabrys}},
  \bibinfo {author} {\bibfnamefont {T.}~\bibnamefont {Oka}}, \bibinfo {author}
  {\bibfnamefont {B.~J.}\ \bibnamefont {Cotterell}}, \bibinfo {author}
  {\bibfnamefont {R.~J.}\ \bibnamefont {Stickland}}, \bibinfo {author}
  {\bibfnamefont {C.}~\bibnamefont {Jungen}},\ and\ \bibinfo {author}
  {\bibfnamefont {A.}~\bibnamefont {Wüest}},\ }\bibfield  {title} {\bibinfo
  {title} {Fine structure of the {H}$_2$ 5g–4f inter-{R}ydberg transition
  revealed by difference frequency laser spectroscopy},\ }\href
  {https://doi.org/10.1063/1.1322634} {\bibfield  {journal} {\bibinfo
  {journal} {J. Chem. Phys.}\ }\textbf {\bibinfo {volume} {113}},\ \bibinfo
  {pages} {10143} (\bibinfo {year} {2000})}\BibitemShut {NoStop}%
\bibitem [{\citenamefont {Beyer}\ \emph {et~al.}(2018)\citenamefont {Beyer},
  \citenamefont {H\"olsch}, \citenamefont {Agner}, \citenamefont {Deiglmayr},
  \citenamefont {Schmutz},\ and\ \citenamefont {Merkt}}]{Beyer2018b}%
  \BibitemOpen
  \bibfield  {author} {\bibinfo {author} {\bibfnamefont {M.}~\bibnamefont
  {Beyer}}, \bibinfo {author} {\bibfnamefont {N.}~\bibnamefont {H\"olsch}},
  \bibinfo {author} {\bibfnamefont {J.~A.}\ \bibnamefont {Agner}}, \bibinfo
  {author} {\bibfnamefont {J.}~\bibnamefont {Deiglmayr}}, \bibinfo {author}
  {\bibfnamefont {H.}~\bibnamefont {Schmutz}},\ and\ \bibinfo {author}
  {\bibfnamefont {F.}~\bibnamefont {Merkt}},\ }\bibfield  {title} {\bibinfo
  {title} {Metrology of high-$n$ {R}ydberg states of molecular hydrogen with
  $\mathrm{\ensuremath{\Delta}}\ensuremath{\nu}/\ensuremath{\nu}=2\ifmmode\times\else\texttimes\fi{}{10}^{\ensuremath{-}10}$
  accuracy},\ }\href {https://link.aps.org/doi/10.1103/PhysRevA.97.012501}
  {\bibfield  {journal} {\bibinfo  {journal} {Phys. Rev. A}\ }\textbf {\bibinfo
  {volume} {97}},\ \bibinfo {pages} {012501} (\bibinfo {year}
  {2018})}\BibitemShut {NoStop}%
\bibitem [{\citenamefont {Gallagher}(1994)}]{Gallagher1994}%
  \BibitemOpen
  \bibfield  {author} {\bibinfo {author} {\bibfnamefont {T.~F.}\ \bibnamefont
  {Gallagher}},\ }\href@noop {} {\emph {\bibinfo {title} {Rydberg Atoms}}}\
  (\bibinfo  {publisher} {Cambridge University Press},\ \bibinfo {year}
  {1994})\BibitemShut {NoStop}%
\bibitem [{\citenamefont {Siłkowski}\ \emph {et~al.}(2021)\citenamefont
  {Siłkowski}, \citenamefont {Zientkiewicz},\ and\ \citenamefont
  {Pachucki}}]{Silkowski2021}%
  \BibitemOpen
  \bibfield  {author} {\bibinfo {author} {\bibfnamefont {M.}~\bibnamefont
  {Siłkowski}}, \bibinfo {author} {\bibfnamefont {M.}~\bibnamefont
  {Zientkiewicz}},\ and\ \bibinfo {author} {\bibfnamefont {K.}~\bibnamefont
  {Pachucki}},\ }\bibfield  {title} {\bibinfo {title} {Accurate
  {Born-Oppenheimer} potentials for excited {$\Sigma^+$} states of the hydrogen
  molecule},\ }\href
  {https://www.sciencedirect.com/science/article/pii/S0065327621000149}
  {\bibfield  {journal} {\bibinfo  {journal} {Adv. Quant. Chem.}\ }\textbf
  {\bibinfo {volume} {83}},\ \bibinfo {pages} {255} (\bibinfo {year}
  {2021})}\BibitemShut {NoStop}%
\bibitem [{\citenamefont {Silkowski}\ and\ \citenamefont
  {Pachucki}()}]{PachuckiPrivate}%
  \BibitemOpen
  \bibfield  {author} {\bibinfo {author} {\bibfnamefont {M.}~\bibnamefont
  {Silkowski}}\ and\ \bibinfo {author} {\bibfnamefont {K.}~\bibnamefont
  {Pachucki}},\ }\href@noop {} {}\bibinfo {howpublished} {private
  communication}\BibitemShut {NoStop}%
\bibitem [{\citenamefont {Jungen}\ \emph {et~al.}(1989)\citenamefont {Jungen},
  \citenamefont {Dabrowski}, \citenamefont {Herzberg},\ and\ \citenamefont
  {Kendall}}]{Jungen1989}%
  \BibitemOpen
  \bibfield  {author} {\bibinfo {author} {\bibfnamefont
  {{\mbox{Ch}}.}~\bibnamefont {Jungen}}, \bibinfo {author} {\bibfnamefont
  {I.}~\bibnamefont {Dabrowski}}, \bibinfo {author} {\bibfnamefont
  {G.}~\bibnamefont {Herzberg}},\ and\ \bibinfo {author} {\bibfnamefont
  {D.~J.~W.}\ \bibnamefont {Kendall}},\ }\bibfield  {title} {\bibinfo {title}
  {High orbital angular momentum states in {H}$_2$ and {D}$_2$. {II}. {T}he
  6h–5g and 6g–5f transitions},\ }\href {https://doi.org/10.1063/1.456824}
  {\bibfield  {journal} {\bibinfo  {journal} {J. Chem. Phys.}\ }\textbf
  {\bibinfo {volume} {91}},\ \bibinfo {pages} {3926} (\bibinfo {year}
  {1989})}\BibitemShut {NoStop}%
\bibitem [{\citenamefont {Yerokhin}\ \emph {et~al.}(2019)\citenamefont
  {Yerokhin}, \citenamefont {Pachucki},\ and\ \citenamefont
  {Patkóš}}]{Yerokhin2019}%
  \BibitemOpen
  \bibfield  {author} {\bibinfo {author} {\bibfnamefont {V.~A.}\ \bibnamefont
  {Yerokhin}}, \bibinfo {author} {\bibfnamefont {K.}~\bibnamefont {Pachucki}},\
  and\ \bibinfo {author} {\bibfnamefont {V.}~\bibnamefont {Patkóš}},\
  }\bibfield  {title} {\bibinfo {title} {Theory of the {L}amb shift in hydrogen
  and light hydrogen-like ions},\ }\href
  {https://onlinelibrary.wiley.com/doi/abs/10.1002/andp.201800324} {\bibfield
  {journal} {\bibinfo  {journal} {Ann. d. Physik}\ }\textbf {\bibinfo {volume}
  {531}},\ \bibinfo {pages} {1800324} (\bibinfo {year} {2019})}\BibitemShut
  {NoStop}%
\bibitem [{\citenamefont {Korobov}\ \emph {et~al.}(2017)\citenamefont
  {Korobov}, \citenamefont {Hilico},\ and\ \citenamefont {Karr}}]{Korobov2017}%
  \BibitemOpen
  \bibfield  {author} {\bibinfo {author} {\bibfnamefont {V.~I.}\ \bibnamefont
  {Korobov}}, \bibinfo {author} {\bibfnamefont {L.}~\bibnamefont {Hilico}},\
  and\ \bibinfo {author} {\bibfnamefont {J.-P.}\ \bibnamefont {Karr}},\
  }\bibfield  {title} {\bibinfo {title} {Fundamental transitions and ionization
  energies of the hydrogen molecular ions with few ppt uncertainty},\ }\href
  {https://link.aps.org/doi/10.1103/PhysRevLett.118.233001} {\bibfield
  {journal} {\bibinfo  {journal} {Phys. Rev. Lett.}\ }\textbf {\bibinfo
  {volume} {118}},\ \bibinfo {pages} {233001} (\bibinfo {year}
  {2017})}\BibitemShut {NoStop}%
\bibitem [{\citenamefont {Korobov}()}]{KorobovPrivate}%
  \BibitemOpen
  \bibfield  {author} {\bibinfo {author} {\bibfnamefont {V.~I.}\ \bibnamefont
  {Korobov}},\ }\href@noop {} {}\bibinfo {howpublished} {private
  communication}\BibitemShut {NoStop}%
\bibitem [{\citenamefont {Le~Roy}\ and\ \citenamefont
  {Barwell}(1975)}]{Leroy1975}%
  \BibitemOpen
  \bibfield  {author} {\bibinfo {author} {\bibfnamefont {R.~J.}\ \bibnamefont
  {Le~Roy}}\ and\ \bibinfo {author} {\bibfnamefont {M.~G.}\ \bibnamefont
  {Barwell}},\ }\bibfield  {title} {\bibinfo {title} {Ground state {D}$_2$
  dissociation energy from the near-dissociation behavior of rotational level
  spacings},\ }\href {https://doi.org/10.1139/p75-248} {\bibfield  {journal}
  {\bibinfo  {journal} {Can. J. Phys.}\ }\textbf {\bibinfo {volume} {53}},\
  \bibinfo {pages} {1983} (\bibinfo {year} {1975})}\BibitemShut {NoStop}%
\bibitem [{\citenamefont {Jungen}\ \emph {et~al.}(1992)\citenamefont {Jungen},
  \citenamefont {Dabrowski}, \citenamefont {Herzberg},\ and\ \citenamefont
  {Vervloet}}]{Jungen1992}%
  \BibitemOpen
  \bibfield  {author} {\bibinfo {author} {\bibfnamefont
  {{\mbox{Ch}}.}~\bibnamefont {Jungen}}, \bibinfo {author} {\bibfnamefont
  {I.}~\bibnamefont {Dabrowski}}, \bibinfo {author} {\bibfnamefont
  {G.}~\bibnamefont {Herzberg}},\ and\ \bibinfo {author} {\bibfnamefont
  {M.}~\bibnamefont {Vervloet}},\ }\bibfield  {title} {\bibinfo {title} {The
  ionization potential of {D}$_2$},\ }\href
  {https://www.sciencedirect.com/science/article/pii/002228529290452T}
  {\bibfield  {journal} {\bibinfo  {journal} {J. Mol. Spectr.}\ }\textbf
  {\bibinfo {volume} {153}},\ \bibinfo {pages} {11} (\bibinfo {year}
  {1992})}\BibitemShut {NoStop}%
\bibitem [{\citenamefont {Ko\l{}os}\ and\ \citenamefont
  {Wolniewicz}(1975)}]{Kolos1975}%
  \BibitemOpen
  \bibfield  {author} {\bibinfo {author} {\bibfnamefont {W.}~\bibnamefont
  {Ko\l{}os}}\ and\ \bibinfo {author} {\bibfnamefont {L.}~\bibnamefont
  {Wolniewicz}},\ }\bibfield  {title} {\bibinfo {title} {Improved potential
  energy curve and vibrational energies for the electronic ground state of the
  hydrogen molecule},\ }\href
  {http://www.sciencedirect.com/science/article/pii/0022285275900831}
  {\bibfield  {journal} {\bibinfo  {journal} {J. Mol. Spectrosc.}\ }\textbf
  {\bibinfo {volume} {54}},\ \bibinfo {pages} {303} (\bibinfo {year}
  {1975})}\BibitemShut {NoStop}%
\bibitem [{\citenamefont {Wolniewicz}(1983)}]{Wolniewicz1983}%
  \BibitemOpen
  \bibfield  {author} {\bibinfo {author} {\bibfnamefont {L.}~\bibnamefont
  {Wolniewicz}},\ }\bibfield  {title} {\bibinfo {title} {The {X$^1\Sigma_g^+$}
  state vibration{-}rotational energies of the {H$_2$}, {HD}, and {D$_2$}
  molecules},\ }\href {https://doi.org/10.1063/1.444580} {\bibfield  {journal}
  {\bibinfo  {journal} {J. Chem. Phys.}\ }\textbf {\bibinfo {volume} {78}},\
  \bibinfo {pages} {6173} (\bibinfo {year} {1983})}\BibitemShut {NoStop}%
\bibitem [{\citenamefont {Ko\l{}os}\ \emph {et~al.}(1986)\citenamefont
  {Ko\l{}os}, \citenamefont {Szalewicz},\ and\ \citenamefont
  {Monkhorst}}]{Kolos1986}%
  \BibitemOpen
  \bibfield  {author} {\bibinfo {author} {\bibfnamefont {W.}~\bibnamefont
  {Ko\l{}os}}, \bibinfo {author} {\bibfnamefont {K.}~\bibnamefont
  {Szalewicz}},\ and\ \bibinfo {author} {\bibfnamefont {H.}~\bibnamefont
  {Monkhorst}},\ }\bibfield  {title} {\bibinfo {title} {New
  {B}orn{-}{O}ppenheimer potential energy curve and vibrational energies for
  the electronic ground state of the hydrogen molecule},\ }\href
  {http://dx.doi.org/10.1063/1.450258} {\bibfield  {journal} {\bibinfo
  {journal} {J. Chem. Phys.}\ }\textbf {\bibinfo {volume} {84}},\ \bibinfo
  {pages} {3278} (\bibinfo {year} {1986})}\BibitemShut {NoStop}%
\bibitem [{\citenamefont {Ko\l{}os}\ and\ \citenamefont
  {Rychlewski}(1993)}]{Kolos1993}%
  \BibitemOpen
  \bibfield  {author} {\bibinfo {author} {\bibfnamefont {W.}~\bibnamefont
  {Ko\l{}os}}\ and\ \bibinfo {author} {\bibfnamefont {J.}~\bibnamefont
  {Rychlewski}},\ }\bibfield  {title} {\bibinfo {title} {Improved theoretical
  dissociation energy and ionization potential for the ground state of the
  hydrogen molecule},\ }\href {http://dx.doi.org/10.1063/1.464023} {\bibfield
  {journal} {\bibinfo  {journal} {J. Chem. Phys.}\ }\textbf {\bibinfo {volume}
  {98}},\ \bibinfo {pages} {3960} (\bibinfo {year} {1993})}\BibitemShut
  {NoStop}%
\bibitem [{\citenamefont {Wolniewicz}(1993)}]{Wolniewicz1993}%
  \BibitemOpen
  \bibfield  {author} {\bibinfo {author} {\bibfnamefont {L.}~\bibnamefont
  {Wolniewicz}},\ }\bibfield  {title} {\bibinfo {title} {Relativistic energies
  of the ground state of the hydrogen molecule},\ }\href
  {http://link.aip.org/link/?JCP/99/1851/1} {\bibfield  {journal} {\bibinfo
  {journal} {\jcp}\ }\textbf {\bibinfo {volume} {99}},\ \bibinfo {pages} {1851}
  (\bibinfo {year} {1993})}\BibitemShut {NoStop}%
\bibitem [{\citenamefont {Wolniewicz}(1995)}]{Wolniewicz1995}%
  \BibitemOpen
  \bibfield  {author} {\bibinfo {author} {\bibfnamefont {L.}~\bibnamefont
  {Wolniewicz}},\ }\bibfield  {title} {\bibinfo {title} {Nonadiabatic energies
  of the ground state of the hydrogen molecule},\ }\href
  {http://link.aip.org/link/?JCP/103/1792/1} {\bibfield  {journal} {\bibinfo
  {journal} {\jcp}\ }\textbf {\bibinfo {volume} {103}},\ \bibinfo {pages}
  {1792} (\bibinfo {year} {1995})}\BibitemShut {NoStop}%
\bibitem [{\citenamefont {Jóźwiak}\ \emph {et~al.}(2020)\citenamefont
  {Jóźwiak}, \citenamefont {Cybulski},\ and\ \citenamefont
  {Wcisło}}]{Jozwiak2020}%
  \BibitemOpen
  \bibfield  {author} {\bibinfo {author} {\bibfnamefont {H.}~\bibnamefont
  {Jóźwiak}}, \bibinfo {author} {\bibfnamefont {H.}~\bibnamefont
  {Cybulski}},\ and\ \bibinfo {author} {\bibfnamefont {P.}~\bibnamefont
  {Wcisło}},\ }\bibfield  {title} {\bibinfo {title} {Hyperfine components of
  all rovibrational quadrupole transitions in the {H}$_2$ and {D}$_2$
  molecules},\ }\href
  {https://doi.org/https://doi.org/10.1016/j.jqsrt.2020.107186} {\bibfield
  {journal} {\bibinfo  {journal} {J. Quant. Spectr. Rad. Transfer}\ }\textbf
  {\bibinfo {volume} {253}},\ \bibinfo {pages} {107186} (\bibinfo {year}
  {2020})}\BibitemShut {NoStop}%
\bibitem [{\citenamefont {Puchalski}\ \emph
  {et~al.}(2019{\natexlab{b}})\citenamefont {Puchalski}, \citenamefont
  {Komasa}, \citenamefont {Czachorowski},\ and\ \citenamefont
  {Pachucki}}]{Puchalski2019b}%
  \BibitemOpen
  \bibfield  {author} {\bibinfo {author} {\bibfnamefont {M.}~\bibnamefont
  {Puchalski}}, \bibinfo {author} {\bibfnamefont {J.}~\bibnamefont {Komasa}},
  \bibinfo {author} {\bibfnamefont {P.}~\bibnamefont {Czachorowski}},\ and\
  \bibinfo {author} {\bibfnamefont {K.}~\bibnamefont {Pachucki}},\ }\bibfield
  {title} {\bibinfo {title} {Nonadiabatic {QED} {Correction} to the
  {Dissociation} {Energy} of the {Hydrogen} {Molecule}},\ }\href
  {https://link.aps.org/doi/10.1103/PhysRevLett.122.103003} {\bibfield
  {journal} {\bibinfo  {journal} {Phys. Rev. Lett.}\ }\textbf {\bibinfo
  {volume} {122}},\ \bibinfo {pages} {103003} (\bibinfo {year}
  {2019}{\natexlab{b}})}\BibitemShut {NoStop}%
\bibitem [{\citenamefont {Tiesinga}\ \emph {et~al.}(2021)\citenamefont
  {Tiesinga}, \citenamefont {Mohr}, \citenamefont {Newell},\ and\ \citenamefont
  {Taylor}}]{CODATA2018}%
  \BibitemOpen
  \bibfield  {author} {\bibinfo {author} {\bibfnamefont {E.}~\bibnamefont
  {Tiesinga}}, \bibinfo {author} {\bibfnamefont {P.~J.}\ \bibnamefont {Mohr}},
  \bibinfo {author} {\bibfnamefont {D.~B.}\ \bibnamefont {Newell}},\ and\
  \bibinfo {author} {\bibfnamefont {B.~N.}\ \bibnamefont {Taylor}},\ }\bibfield
   {title} {\bibinfo {title} {{CODATA Recommended Values of the Fundamental
  Physical Constants: 2018}},\ }\href {https://doi.org/10.1063/5.0064853}
  {\bibfield  {journal} {\bibinfo  {journal} {J. Phys. Chem. Ref. Data}\
  }\textbf {\bibinfo {volume} {50}},\ \bibinfo {pages} {033105} (\bibinfo
  {year} {2021})}\BibitemShut {NoStop}%
\bibitem [{\citenamefont {Pohl}\ \emph {et~al.}(2016)\citenamefont {Pohl},
  \citenamefont {Nez}, \citenamefont {Fernandes}, \citenamefont {Amaro},
  \citenamefont {Biraben}, \citenamefont {Cardoso}, \citenamefont {Covita},
  \citenamefont {Dax}, \citenamefont {Dhawan}, \citenamefont {Diepold},
  \citenamefont {Giesen}, \citenamefont {Gouvea}, \citenamefont {Graf},
  \citenamefont {Hänsch}, \citenamefont {Indelicato}, \citenamefont {Julien},
  \citenamefont {Knowles}, \citenamefont {Kottmann}, \citenamefont {Bigot},
  \citenamefont {Liu}, \citenamefont {Lopes}, \citenamefont {Ludhova},
  \citenamefont {Monteiro}, \citenamefont {Mulhauser}, \citenamefont {Nebel},
  \citenamefont {Rabinowitz}, \citenamefont {dos Santos}, \citenamefont
  {Schaller}, \citenamefont {Schuhmann}, \citenamefont {Schwob}, \citenamefont
  {Taqqu}, \citenamefont {Veloso},\ and\ \citenamefont {Antognini}}]{Pohl2016}%
  \BibitemOpen
  \bibfield  {author} {\bibinfo {author} {\bibfnamefont {R.}~\bibnamefont
  {Pohl}}, \bibinfo {author} {\bibfnamefont {F.}~\bibnamefont {Nez}}, \bibinfo
  {author} {\bibfnamefont {L.~M.~P.}\ \bibnamefont {Fernandes}}, \bibinfo
  {author} {\bibfnamefont {F.~D.}\ \bibnamefont {Amaro}}, \bibinfo {author}
  {\bibfnamefont {F.}~\bibnamefont {Biraben}}, \bibinfo {author} {\bibfnamefont
  {J.~M.~R.}\ \bibnamefont {Cardoso}}, \bibinfo {author} {\bibfnamefont
  {D.~S.}\ \bibnamefont {Covita}}, \bibinfo {author} {\bibfnamefont
  {A.}~\bibnamefont {Dax}}, \bibinfo {author} {\bibfnamefont {S.}~\bibnamefont
  {Dhawan}}, \bibinfo {author} {\bibfnamefont {M.}~\bibnamefont {Diepold}},
  \bibinfo {author} {\bibfnamefont {A.}~\bibnamefont {Giesen}}, \bibinfo
  {author} {\bibfnamefont {A.~L.}\ \bibnamefont {Gouvea}}, \bibinfo {author}
  {\bibfnamefont {T.}~\bibnamefont {Graf}}, \bibinfo {author} {\bibfnamefont
  {T.~W.}\ \bibnamefont {Hänsch}}, \bibinfo {author} {\bibfnamefont
  {P.}~\bibnamefont {Indelicato}}, \bibinfo {author} {\bibfnamefont
  {L.}~\bibnamefont {Julien}}, \bibinfo {author} {\bibfnamefont
  {P.}~\bibnamefont {Knowles}}, \bibinfo {author} {\bibfnamefont
  {F.}~\bibnamefont {Kottmann}}, \bibinfo {author} {\bibfnamefont {E.-O.~L.}\
  \bibnamefont {Bigot}}, \bibinfo {author} {\bibfnamefont {Y.-W.}\ \bibnamefont
  {Liu}}, \bibinfo {author} {\bibfnamefont {J.~A.~M.}\ \bibnamefont {Lopes}},
  \bibinfo {author} {\bibfnamefont {L.}~\bibnamefont {Ludhova}}, \bibinfo
  {author} {\bibfnamefont {C.~M.~B.}\ \bibnamefont {Monteiro}}, \bibinfo
  {author} {\bibfnamefont {F.}~\bibnamefont {Mulhauser}}, \bibinfo {author}
  {\bibfnamefont {T.}~\bibnamefont {Nebel}}, \bibinfo {author} {\bibfnamefont
  {P.}~\bibnamefont {Rabinowitz}}, \bibinfo {author} {\bibfnamefont {J.~M.~F.}\
  \bibnamefont {dos Santos}}, \bibinfo {author} {\bibfnamefont {L.~A.}\
  \bibnamefont {Schaller}}, \bibinfo {author} {\bibfnamefont {K.}~\bibnamefont
  {Schuhmann}}, \bibinfo {author} {\bibfnamefont {C.}~\bibnamefont {Schwob}},
  \bibinfo {author} {\bibfnamefont {D.}~\bibnamefont {Taqqu}}, \bibinfo
  {author} {\bibfnamefont {J.~F. C.~A.}\ \bibnamefont {Veloso}},\ and\ \bibinfo
  {author} {\bibfnamefont {A.}~\bibnamefont {Antognini}},\ }\bibfield  {title}
  {\bibinfo {title} {Laser spectroscopy of muonic deuterium},\ }\href
  {https://www.science.org/doi/abs/10.1126/science.aaf2468} {\bibfield
  {journal} {\bibinfo  {journal} {Science}\ }\textbf {\bibinfo {volume}
  {353}},\ \bibinfo {pages} {669} (\bibinfo {year} {2016})}\BibitemShut
  {NoStop}%
\bibitem [{\citenamefont {Parthey}\ \emph {et~al.}(2011)\citenamefont
  {Parthey}, \citenamefont {Matveev}, \citenamefont {Alnis}, \citenamefont
  {Bernhardt}, \citenamefont {Beyer}, \citenamefont {Holzwarth}, \citenamefont
  {Maistrou}, \citenamefont {Pohl}, \citenamefont {Predehl}, \citenamefont
  {Udem}, \citenamefont {Wilken}, \citenamefont {Kolachevsky}, \citenamefont
  {Abgrall}, \citenamefont {Rovera}, \citenamefont {Salomon}, \citenamefont
  {Laurent},\ and\ \citenamefont {H\"ansch}}]{Parthey2011}%
  \BibitemOpen
  \bibfield  {author} {\bibinfo {author} {\bibfnamefont {C.~G.}\ \bibnamefont
  {Parthey}}, \bibinfo {author} {\bibfnamefont {A.}~\bibnamefont {Matveev}},
  \bibinfo {author} {\bibfnamefont {J.}~\bibnamefont {Alnis}}, \bibinfo
  {author} {\bibfnamefont {B.}~\bibnamefont {Bernhardt}}, \bibinfo {author}
  {\bibfnamefont {A.}~\bibnamefont {Beyer}}, \bibinfo {author} {\bibfnamefont
  {R.}~\bibnamefont {Holzwarth}}, \bibinfo {author} {\bibfnamefont
  {A.}~\bibnamefont {Maistrou}}, \bibinfo {author} {\bibfnamefont
  {R.}~\bibnamefont {Pohl}}, \bibinfo {author} {\bibfnamefont {K.}~\bibnamefont
  {Predehl}}, \bibinfo {author} {\bibfnamefont {T.}~\bibnamefont {Udem}},
  \bibinfo {author} {\bibfnamefont {T.}~\bibnamefont {Wilken}}, \bibinfo
  {author} {\bibfnamefont {N.}~\bibnamefont {Kolachevsky}}, \bibinfo {author}
  {\bibfnamefont {M.}~\bibnamefont {Abgrall}}, \bibinfo {author} {\bibfnamefont
  {D.}~\bibnamefont {Rovera}}, \bibinfo {author} {\bibfnamefont
  {C.}~\bibnamefont {Salomon}}, \bibinfo {author} {\bibfnamefont
  {P.}~\bibnamefont {Laurent}},\ and\ \bibinfo {author} {\bibfnamefont {T.~W.}\
  \bibnamefont {H\"ansch}},\ }\bibfield  {title} {\bibinfo {title} {Improved
  measurement of the hydrogen {1S\ensuremath{-}2S} transition frequency},\
  }\href {https://link.aps.org/doi/10.1103/PhysRevLett.107.203001} {\bibfield
  {journal} {\bibinfo  {journal} {Phys. Rev. Lett.}\ }\textbf {\bibinfo
  {volume} {107}},\ \bibinfo {pages} {203001} (\bibinfo {year}
  {2011})}\BibitemShut {NoStop}%
\bibitem [{\citenamefont {Parthey}\ \emph {et~al.}(2010)\citenamefont
  {Parthey}, \citenamefont {Matveev}, \citenamefont {Alnis}, \citenamefont
  {Pohl}, \citenamefont {Udem}, \citenamefont {Jentschura}, \citenamefont
  {Kolachevsky},\ and\ \citenamefont {H\"ansch}}]{Parthey2010}%
  \BibitemOpen
  \bibfield  {author} {\bibinfo {author} {\bibfnamefont {C.~G.}\ \bibnamefont
  {Parthey}}, \bibinfo {author} {\bibfnamefont {A.}~\bibnamefont {Matveev}},
  \bibinfo {author} {\bibfnamefont {J.}~\bibnamefont {Alnis}}, \bibinfo
  {author} {\bibfnamefont {R.}~\bibnamefont {Pohl}}, \bibinfo {author}
  {\bibfnamefont {T.}~\bibnamefont {Udem}}, \bibinfo {author} {\bibfnamefont
  {U.~D.}\ \bibnamefont {Jentschura}}, \bibinfo {author} {\bibfnamefont
  {N.}~\bibnamefont {Kolachevsky}},\ and\ \bibinfo {author} {\bibfnamefont
  {T.~W.}\ \bibnamefont {H\"ansch}},\ }\bibfield  {title} {\bibinfo {title}
  {Precision measurement of the hydrogen-deuterium {1S\ensuremath{-}2S} isotope
  shift},\ }\href {https://link.aps.org/doi/10.1103/PhysRevLett.104.233001}
  {\bibfield  {journal} {\bibinfo  {journal} {Phys. Rev. Lett.}\ }\textbf
  {\bibinfo {volume} {104}},\ \bibinfo {pages} {233001} (\bibinfo {year}
  {2010})}\BibitemShut {NoStop}%
\bibitem [{\citenamefont {Alighanbari}\ \emph {et~al.}(2020)\citenamefont
  {Alighanbari}, \citenamefont {Giri}, \citenamefont {Constantin},
  \citenamefont {Korobov},\ and\ \citenamefont {Schiller}}]{Alighanbari2020}%
  \BibitemOpen
  \bibfield  {author} {\bibinfo {author} {\bibfnamefont {S.}~\bibnamefont
  {Alighanbari}}, \bibinfo {author} {\bibfnamefont {G.~S.}\ \bibnamefont
  {Giri}}, \bibinfo {author} {\bibfnamefont {F.~L.}\ \bibnamefont
  {Constantin}}, \bibinfo {author} {\bibfnamefont {V.~I.}\ \bibnamefont
  {Korobov}},\ and\ \bibinfo {author} {\bibfnamefont {S.}~\bibnamefont
  {Schiller}},\ }\bibfield  {title} {\bibinfo {title} {Precise test of quantum
  electrodynamics and determination of fundamental constants with {HD}$^+$
  ions},\ }\href {https://doi.org/10.1038/s41586-020-2261-5} {\bibfield
  {journal} {\bibinfo  {journal} {Nature}\ }\textbf {\bibinfo {volume} {581}},\
  \bibinfo {pages} {152–158} (\bibinfo {year} {2020})}\BibitemShut {NoStop}%
\bibitem [{\citenamefont {Patra}\ \emph {et~al.}(2020)\citenamefont {Patra},
  \citenamefont {Germann}, \citenamefont {Karr}, \citenamefont {Haidar},
  \citenamefont {Hilico}, \citenamefont {Korobov}, \citenamefont {Cozijn},
  \citenamefont {Eikema}, \citenamefont {Ubachs},\ and\ \citenamefont
  {Koelemeij}}]{Patra2020}%
  \BibitemOpen
  \bibfield  {author} {\bibinfo {author} {\bibfnamefont {S.}~\bibnamefont
  {Patra}}, \bibinfo {author} {\bibfnamefont {M.}~\bibnamefont {Germann}},
  \bibinfo {author} {\bibfnamefont {J.-P.}\ \bibnamefont {Karr}}, \bibinfo
  {author} {\bibfnamefont {M.}~\bibnamefont {Haidar}}, \bibinfo {author}
  {\bibfnamefont {L.}~\bibnamefont {Hilico}}, \bibinfo {author} {\bibfnamefont
  {V.~I.}\ \bibnamefont {Korobov}}, \bibinfo {author} {\bibfnamefont
  {F.~M.~J.}\ \bibnamefont {Cozijn}}, \bibinfo {author} {\bibfnamefont
  {K.~S.~E.}\ \bibnamefont {Eikema}}, \bibinfo {author} {\bibfnamefont
  {W.}~\bibnamefont {Ubachs}},\ and\ \bibinfo {author} {\bibfnamefont
  {J.~C.~J.}\ \bibnamefont {Koelemeij}},\ }\bibfield  {title} {\bibinfo {title}
  {Proton-electron mass ratio from laser spectroscopy of {HD}$^+$ at the
  part-per-trillion level},\ }\href {https://doi.org/10.1126/science.aba0453}
  {\bibfield  {journal} {\bibinfo  {journal} {Science}\ }\textbf {\bibinfo
  {volume} {369}},\ \bibinfo {pages} {1238} (\bibinfo {year}
  {2020})}\BibitemShut {NoStop}%
\bibitem [{\citenamefont {Schmidt}\ \emph {et~al.}(2020)\citenamefont
  {Schmidt}, \citenamefont {Louvradoux}, \citenamefont {Heinrich},
  \citenamefont {Sillitoe}, \citenamefont {Simpson}, \citenamefont {Karr},\
  and\ \citenamefont {Hilico}}]{Schmidt2020}%
  \BibitemOpen
  \bibfield  {author} {\bibinfo {author} {\bibfnamefont {J.}~\bibnamefont
  {Schmidt}}, \bibinfo {author} {\bibfnamefont {T.}~\bibnamefont {Louvradoux}},
  \bibinfo {author} {\bibfnamefont {J.}~\bibnamefont {Heinrich}}, \bibinfo
  {author} {\bibfnamefont {N.}~\bibnamefont {Sillitoe}}, \bibinfo {author}
  {\bibfnamefont {M.}~\bibnamefont {Simpson}}, \bibinfo {author} {\bibfnamefont
  {J.-P.}\ \bibnamefont {Karr}},\ and\ \bibinfo {author} {\bibfnamefont
  {L.}~\bibnamefont {Hilico}},\ }\bibfield  {title} {\bibinfo {title}
  {Trapping, cooling, and photodissociation analysis of state-selected
  {H}$_{2}^{+}$ ions produced by ($3+1$) multiphoton ionization},\ }\href
  {https://link.aps.org/doi/10.1103/PhysRevApplied.14.024053} {\bibfield
  {journal} {\bibinfo  {journal} {Phys. Rev. Appl.}\ }\textbf {\bibinfo
  {volume} {14}},\ \bibinfo {pages} {024053} (\bibinfo {year}
  {2020})}\BibitemShut {NoStop}%
\bibitem [{\citenamefont {Pohl}\ \emph {et~al.}(2010)\citenamefont {Pohl},
  \citenamefont {Antognini}, \citenamefont {Nez}, \citenamefont {Amaro},
  \citenamefont {Biraben}, \citenamefont {Cardoso}, \citenamefont {Covita},
  \citenamefont {Dax}, \citenamefont {Dhawan}, \citenamefont {Fernandes},
  \citenamefont {Giesen}, \citenamefont {Graf}, \citenamefont {H{\"{a}}nsch},
  \citenamefont {Indelicato}, \citenamefont {Julien}, \citenamefont {Kao},
  \citenamefont {Knowles}, \citenamefont {Bigot}, \citenamefont {Liu},
  \citenamefont {Lopes}, \citenamefont {Ludhova}, \citenamefont {Monteiro},
  \citenamefont {Mulhauser}, \citenamefont {Nebel}, \citenamefont {Rabinowitz},
  \citenamefont {dos Santos}, \citenamefont {Schaller}, \citenamefont
  {Schuhmann}, \citenamefont {Schwob}, \citenamefont {Taqqu}, \citenamefont
  {Veloso},\ and\ \citenamefont {Kottmann}}]{Pohl2010}%
  \BibitemOpen
  \bibfield  {author} {\bibinfo {author} {\bibfnamefont {R.}~\bibnamefont
  {Pohl}}, \bibinfo {author} {\bibfnamefont {A.}~\bibnamefont {Antognini}},
  \bibinfo {author} {\bibfnamefont {F.}~\bibnamefont {Nez}}, \bibinfo {author}
  {\bibfnamefont {F.~D.}\ \bibnamefont {Amaro}}, \bibinfo {author}
  {\bibfnamefont {F.}~\bibnamefont {Biraben}}, \bibinfo {author} {\bibfnamefont
  {J.~M.~R.}\ \bibnamefont {Cardoso}}, \bibinfo {author} {\bibfnamefont
  {D.~S.}\ \bibnamefont {Covita}}, \bibinfo {author} {\bibfnamefont
  {A.}~\bibnamefont {Dax}}, \bibinfo {author} {\bibfnamefont {S.}~\bibnamefont
  {Dhawan}}, \bibinfo {author} {\bibfnamefont {L.~M.~P.}\ \bibnamefont
  {Fernandes}}, \bibinfo {author} {\bibfnamefont {A.}~\bibnamefont {Giesen}},
  \bibinfo {author} {\bibfnamefont {T.}~\bibnamefont {Graf}}, \bibinfo {author}
  {\bibfnamefont {T.~W.}\ \bibnamefont {H{\"{a}}nsch}}, \bibinfo {author}
  {\bibfnamefont {P.}~\bibnamefont {Indelicato}}, \bibinfo {author}
  {\bibfnamefont {L.}~\bibnamefont {Julien}}, \bibinfo {author} {\bibfnamefont
  {C.~Y.}\ \bibnamefont {Kao}}, \bibinfo {author} {\bibfnamefont
  {P.}~\bibnamefont {Knowles}}, \bibinfo {author} {\bibfnamefont {E.-O.~L.}\
  \bibnamefont {Bigot}}, \bibinfo {author} {\bibfnamefont {Y.-W.}\ \bibnamefont
  {Liu}}, \bibinfo {author} {\bibfnamefont {J.~A.~M.}\ \bibnamefont {Lopes}},
  \bibinfo {author} {\bibfnamefont {L.}~\bibnamefont {Ludhova}}, \bibinfo
  {author} {\bibfnamefont {C.~M.~B.}\ \bibnamefont {Monteiro}}, \bibinfo
  {author} {\bibfnamefont {F.}~\bibnamefont {Mulhauser}}, \bibinfo {author}
  {\bibfnamefont {T.}~\bibnamefont {Nebel}}, \bibinfo {author} {\bibfnamefont
  {P.}~\bibnamefont {Rabinowitz}}, \bibinfo {author} {\bibfnamefont {J.~M.~F.}\
  \bibnamefont {dos Santos}}, \bibinfo {author} {\bibfnamefont {L.~A.}\
  \bibnamefont {Schaller}}, \bibinfo {author} {\bibfnamefont {K.}~\bibnamefont
  {Schuhmann}}, \bibinfo {author} {\bibfnamefont {C.}~\bibnamefont {Schwob}},
  \bibinfo {author} {\bibfnamefont {D.}~\bibnamefont {Taqqu}}, \bibinfo
  {author} {\bibfnamefont {J.~F. C.~A.}\ \bibnamefont {Veloso}},\ and\ \bibinfo
  {author} {\bibfnamefont {F.}~\bibnamefont {Kottmann}},\ }\bibfield  {title}
  {\bibinfo {title} {The size of the proton},\ }\href
  {http://www.nature.com/nature/journal/v466/n7303/full/nature09250.html}
  {\bibfield  {journal} {\bibinfo  {journal} {Nature}\ }\textbf {\bibinfo
  {volume} {466}},\ \bibinfo {pages} {213} (\bibinfo {year}
  {2010})}\BibitemShut {NoStop}%
\end{thebibliography}

%

\end{document}